\documentstyle[epsfig]{mn2e}

\begin{document}

\title[WiggleZ survey: selection function]{The WiggleZ Dark Energy
  Survey: the selection function and $z=0.6$ galaxy power spectrum}

\author[Blake et al.]{\parbox[t]{\textwidth}{Chris
    Blake$^1$\footnotemark, Sarah Brough$^1$, Matthew Colless$^2$,
    Warrick Couch$^1$, Scott Croom$^3$, Tamara Davis$^{4,5}$, Michael
    J.\ Drinkwater$^4$, Karl Forster$^6$, Karl Glazebrook$^1$, \\ Ben
    Jelliffe$^3$, Russell J.\ Jurek$^4$, I-hui Li$^1$, Barry
    Madore$^7$, Chris Martin$^6$, \\ Kevin Pimbblet$^8$, Gregory
    B.\ Poole$^1$, Michael Pracy$^1$, Rob Sharp$^2$, \\ Emily
    Wisnioski$^1$, David Woods$^9$ and Ted Wyder$^6$} \\ \\ $^1$
  Centre for Astrophysics \& Supercomputing, Swinburne University of
  Technology, P.O. Box 218, Hawthorn, VIC 3122, Australia \\ $^2$
  Anglo-Australian Observatory, P.O. Box 296, Epping, NSW 2121,
  Australia \\ $^3$ School of Physics, University of Sydney, NSW 2006,
  Australia \\ $^4$ Department of Physics, University of Queensland,
  Brisbane, QLD 4072, Australia \\ $^5$ Dark Cosmology Centre, Niels
  Bohr Institute, University of Copenhagen, Juliane Maries Vej 30,
  DK-2100 Copenhagen, Denmark \\ $^6$ California Institute of
  Technology, MC 405-47, 1200 East California Boulevard, Pasadena, CA
  91125, United States \\ $^7$ Observatories of the Carnegie Institute
  of Washington, 813 Santa Barbara St., Pasadena, CA 91101, United
  States \\ $^8$ School of Physics, Monash University, Clayton, VIC
  3800, Australia \\ $^9$ Department of Physics \& Astronomy,
  University of British Columbia, 6224 Agricultural Road, Vancouver,
  B.C., V6T 1Z1, Canada}

\maketitle

\begin{abstract}
  We report one of the most accurate measurements of the
  three-dimensional large-scale galaxy power spectrum achieved to
  date, using $56{,}159$ redshifts of bright emission-line galaxies at
  effective redshift $z \approx 0.6$ from the WiggleZ Dark Energy
  Survey at the Anglo-Australian Telescope.  We describe in detail how
  we construct the survey selection function allowing for the varying
  target completeness and redshift completeness.  We measure the total
  power with an accuracy of approximately $5\%$ in wavenumber bands of
  $\Delta k = 0.01 \, h$ Mpc$^{-1}$.  A model power spectrum including
  non-linear corrections, combined with a linear galaxy bias factor
  and a simple model for redshift-space distortions, provides a good
  fit to our data for scales $k < 0.4 \, h$ Mpc$^{-1}$.  The
  large-scale shape of the power spectrum is consistent with the
  best-fitting matter and baryon densities determined by observations
  of the Cosmic Microwave Background radiation.  By splitting the
  power spectrum measurement as a function of tangential and radial
  wavenumbers we delineate the characteristic imprint of peculiar
  velocities.  We use these to determine the growth rate of structure
  as a function of redshift in the range $0.4 < z < 0.8$, including a
  data point at $z=0.78$ with an accuracy of $20\%$.  Our growth rate
  measurements are a close match to the self-consistent prediction of
  the $\Lambda$CDM model.  The WiggleZ Survey data will allow a wide
  range of investigations into the cosmological model, cosmic
  expansion and growth history, topology of cosmic structure, and
  Gaussianity of the initial conditions.  Our calculation of the
  survey selection function will be released at a future date via our
  website {\tt wigglez.swin.edu.au}.
\end{abstract}
\begin{keywords}
surveys, large-scale structure of Universe, cosmological parameters
\end{keywords}

\section{Introduction}
\renewcommand{\thefootnote}{\fnsymbol{footnote}}
\setcounter{footnote}{1}
\footnotetext{E-mail: cblake@astro.swin.edu.au}

The pattern of density fluctuations in the low-redshift Universe
results from the physical processes which govern the evolution of
matter perturbations after the Big Bang.  In the early Universe, the
primordial spectrum of fluctuations created by inflation is processed
before recombination in a manner depending on the physical matter
density, baryon fraction and massive neutrino fraction (e.g.\ Bond \&
Efstathiou 1984; Bardeen et al.\ 1986; Holtzman 1989; Hu \& Sugiyama
1996; Eisenstein \& Hu 1998).  After recombination, perturbations of
all scales are amplified by gravity at an identical rate whilst linear
theory applies.  This growth rate depends on the matter and dark
energy components which drive the cosmic expansion (e.g.\ Heath 1977;
Hamilton 2001; Linder \& Jenkins 2003; Percival 2005).  The growth of
fluctuations enters a non-linear regime at progressively larger scales
at lower redshifts: in today's Universe, only perturbations with
Fourier wavescales $k < 0.1 \, h$ Mpc$^{-1}$ evolve linearly to a good
approximation (e.g.\ Smith et al.\ 2003; Jeong \& Komatsu 2006;
McDonald 2007).

The clustering pattern of galaxies at different redshifts is related
to the underlying density fluctuations and may be used to test this
model of structure formation.  The shape of the clustering power
spectrum -- the relative amplitudes of large-scale and small-scale
modes -- depends on the composition of the early Universe and may be
used to extract information about the matter and baryon fractions
(e.g.\ Tegmark et al.\ 2004a; Cole et al.\ 2005; Percival et al.\
2007b).  The amplitude of the clustering power spectrum as a function
of redshift, together with the pattern of redshift-space distortions
induced by galaxy peculiar velocities, can be used to measure the
growth rate of structure (e.g.\ Hamilton 1992; Hawkins et al.\ 2003;
Guzzo et al.\ 2008; Percival \& White 2009).  Higher-order or
topological descriptors of the density field, such as the bispectrum
or genus, can be applied to test whether the initial conditions are
consistent with scale-invariant Gaussian random perturbations
generated by inflation (e.g.\ Gott, Dickinson \& Melott 1986; Fry \&
Scherrer 1994; Sefusatti \& Komatsu 2007; James, Lewis \& Colless
2007).

The interpretation of the shape and amplitude of the galaxy power
spectrum is complicated by several factors.  Firstly, the manner in
which galaxies trace the density field -- the ``galaxy bias'' -- is in
general a complex function of scale, dark matter halo mass, galaxy
type and redshift (Dekel \& Lahav 1999; Tegmark \& Bromley 1999; Wild
et al.\ 2005; Conway et al.\ 2005; Percival et al.\ 2007a; Smith,
Scoccimarro \& Sheth 2007; Cresswell \& Percival 2009). However, the
bias of galaxy fluctuations on sufficiently large scales ($k < 0.1 \,
h$ Mpc$^{-1}$ at $z=0$) appears to be well-described by a simple
constant of proportionality whose value depends on galaxy type and
luminosity, or more fundamentally dark matter halo mass (Peacock \&
Dodds 1994; Scherrer \& Weinberg 1998; Verde et al.\ 2002).  Secondly,
small-scale density perturbations eventually begin to evolve in a
non-linear fashion requiring more complex modelling techniques such as
higher-order perturbation theory or numerical $N$-body simulations
(Smith et al.\ 2003; Jeong \& Komatsu 2006; McDonald 2007).  Thirdly,
there is a practical challenge of acquiring galaxy survey data across
a ``fair sample'' of the Universe (Tegmark 1997).  For the large-scale
linear modes of clustering, which provide the most robust link to
underlying theory, this sample must map a volume of the order 1
Gpc$^3$ using of the order $10^5$ galaxies.  These demands require
multi-year campaigns with ground-based telescopes utilizing hundreds
of clear nights (Glazebrook \& Blake 2005).

Despite these challenges, a series of galaxy redshift surveys have
been undertaken to provide such datasets at redshifts $z < 0.5$.  The
state-of-the-art projects which have mapped the ``local'' ($z \approx
0.1$) Universe are the 2-degree Field Galaxy Redshift Survey (2dFGRS;
Colless et al.\ 2001) and the Sloan Digital Sky Survey (SDSS; York et
al.\ 2000).  The 2dFGRS obtained redshifts for $2 \times 10^5$
galaxies covering 1500 deg$^2$ in the period between 1997 and 2002.
The ``main'' spectroscopic survey of the SDSS gathered $8 \times 10^5$
galaxy redshifts over 8000 deg$^2$ between the years 2000 and 2005.
The SDSS project also included observations of $1 \times 10^5$
Luminous Red Galaxies (LRGs) reaching up to a redshift $z = 0.5$
(Eisenstein et al.\ 2001).

These datasets have provided a rich source of information about the
clustering of galaxies.  For example, power spectra have been
extracted for the 2dFGRS by Percival et al.\ (2001) and Cole et al.\
(2005); for the SDSS ``main'' galaxy sample by Pope et al.\ (2004),
Tegmark et al.\ (2004a) and Percival et al.\ (2007a); and for the LRGs
by Eisenstein et al.\ (2005), Huetsi (2006), Tegmark et al.\ (2006)
and Percival et al.\ (2007b).  Analysis of these surveys, in
combination with the Cosmic Microwave Background fluctuations, has
confirmed that we inhabit a low-density Universe where matter today
provides only $25-30\%$ of the total energy governing the large-scale
dynamics, with the rest located in a mysterious ``dark energy''
component.  In addition the baryonic fraction of the matter is only
$15-20\%$, with the remainder composed of non-baryonic, cold particles
whose nature is currently unknown (e.g.\ Percival et al.\ 2002;
Tegmark et al.\ 2004b; Tegmark et al.\ 2006; Komatsu et al.\ 2009).
The clustering pattern is also sensitive to the presence of hot dark
matter such as massive neutrinos, which comprise a small fraction of
the energy budget (Elgaroy et al.\ 2002; Seljak et al.\ 2005).

These galaxy surveys also describe how the underlying density
fluctuations are modulated by galaxy bias (Verde et al.\ 2002; Wild et
al.\ 2005; Conway et al.\ 2005; Percival et al.\ 2007a; Cresswell \&
Percival 2009).  In this context the comparison of power spectrum
measurements from the 2dFGRS and SDSS, which targeted galaxy
populations selected in blue and red optical wavebands respectively,
is of particular interest.  When the differing galaxy types in these
surveys are assigned linear bias factors, the resulting model fits to
the linear-regime power spectra produce best-fitting matter densities
which are inconsistent at the statistical level of $2\sigma$.  Careful
treatment of scale-dependent and luminosity-dependent galaxy bias can
potentially explain this discrepancy (Percival et al.\ 2007a; Sanchez
\& Cole 2008).

There are strong motivations for extending these large-scale structure
measurements to higher redshifts ($z > 0.5$).  Firstly, the growth of
structure implies that the linear regime of evolving perturbations
extends to smaller scales at higher redshifts, enabling cleaner and
more accurate model fits.  Secondly, the shape of the survey cone
allows access to significantly greater cosmic volumes at higher
redshift, enabling more accurate determinations of the large-scale
power spectrum amplitude.  Thirdly, baryon oscillations in galaxy
power spectra at different redshifts may be used as a standard ruler
to extract the cosmic distance-redshift relation and infer the
properties of dark energy (Blake \& Glazebrook 2003; Seo \& Eisenstein
2003; Hu \& Haiman 2003).  Fourthly, measurements of the growth of
cosmic structure as a function of redshift increases our ability to
discriminate between dark energy models including modifications to
Einstein's theory of gravity (Guzzo et al.\ 2008; Wang 2008; White,
Song \& Percival 2009).

Our current tools for probing the matter power spectrum at redshifts
$z > 0.5$ are limited.  The clustering of high-redshift quasars has
been studied by the 2dF Quasar Survey (Outram et al.\ 2003) and the
SDSS (Ross et al.\ 2009) but the scarcity of QSOs implies that the
large-scale clustering measurements are strongly limited by shot
noise.  Photometric redshifts from imaging surveys have been used to
study the projected clustering pattern in redshift slices (Blake et
al.\ 2007; Padmanabhan et al.\ 2007).  However, this approach loses
the information from small-scale radial clustering modes (Blake \&
Bridle 2005) and in particular prevents the extraction of the patterns
of peculiar velocities, which are swamped by photometric redshift
errors.  Alternatively, fluctuations in the Lyman-$\alpha$ forest
absorption spectrum on the sight lines to bright quasars have been
used to infer the amplitude of small-scale clustering fluctuations in
the high-redshift Universe (Croft et al.\ 2002; McDonald et al.\ 2005;
McDonald et al.\ 2006).  However, this method is potentially
susceptible to systematic modelling errors and is only applicable at
redshifts $z > 2.3$ where the Lyman-$\alpha$ absorption lines pass
into optical wavebands.

The WiggleZ Dark Energy Survey at the Anglo-Australian Telescope
(Drinkwater et al.\ 2010) will provide the next step forwards in
large-scale spectroscopic galaxy redshift surveys, mapping a cosmic
volume of the order 1 Gpc$^3$ over the redshift range $z < 1$.  The
survey, which began in August 2006 and is scheduled to finish in July
2010, is obtaining of the order $200{,}000$ redshifts for UV-selected
emission-line galaxies covering of the order 1000 deg$^2$ of
equatorial sky.  The principal scientific goal is to measure baryon
oscillations in the galaxy power spectrum in redshift bins up to $z =
1$ and provide a robust measurement of the dark energy model.  The
dataset will also trace the density field over unprecedented cosmic
volumes at $z > 0.5$, providing a sample comparable to the SDSS LRG
catalogue at $z < 0.5$.  Moreover, the spatial overlap between the
WiggleZ and LRG catalogues in the redshift range $0.3 < z < 0.5$ will
allow careful studies of the systematic effects of galaxy bias on
power spectrum estimation.

This paper presents a determination of the current WiggleZ survey
selection function and galaxy power spectrum, using a dataset
comprising of the order $25\%$ of the final survey observations.  The
selection function, which describes the angular and radial survey
coverage in the absence of clustering, is complicated by the
relatively high level of incompleteness in the survey affecting both
the parent target catalogues and the spectroscopic follow-up
observations (although this latter type of incompleteness will
decrease as the survey progresses).  However, we demonstrate that
despite these complications the galaxy clustering power spectrum may
be successfully extracted and already provides accurate tests of the
cosmological model that rival lower-redshift surveys.

The structure of this paper is as follows: in Section \ref{secsel} we
present a detailed account of the survey selection function including
the coverage masks, completeness of our UV imaging catalogues,
variations in redshift completeness, redshift distribution as a
function of sky position, and redshift blunder rate.  In Section
\ref{secpowerspec} we describe our power spectrum calculation and its
correction for redshift blunders.  We compare the predictions of
cosmological models to the resulting power spectra in Section
\ref{secparfit} and itemize our conclusions in Section \ref{secconc}.
We note that we construct our selection function for a fiducial flat
$\Lambda$CDM cosmological model with matter density $\Omega_{\rm m} =
0.3$.

\section{WiggleZ survey selection function}
\label{secsel}

An overview of the WiggleZ Survey observing strategy and galaxy
selection criteria is presented by Blake et al.\ (2009) and Drinkwater
et al.\ (2010) and summarized in Table \ref{tabselection}.  Briefly,
targets are chosen by a joint selection in UV and optical wavebands,
using observations by the Galaxy Evolution Explorer satellite (GALEX)
matched with ground-based optical imaging.  A series of magnitude and
colour cuts is used to preferentially select high-redshift
star-forming galaxies with bright emission lines, which are observed
using the AAOmega multi-object spectrograph at the Anglo-Australian
Telescope (AAT).

\begin{table*}
\caption{Summary of the WiggleZ Survey galaxy selection criteria.}
\label{tabselection}
\begin{tabular}{cc}
\hline
UV magnitude cut & $NUV < 22.8$ \\
Optical magnitude cut & $20 < r < 22.5$ \\
UV colour cut & $FUV$ dropout or $FUV-NUV > 1$ \\
Optical-UV colour cut & $-0.5 < NUV-r < 2$ \\
Optical colour cut & If $g < 22.5$ and $i < 21.5$, then $(r-i) < (g-r-0.1)$ and $(r-i) < 0.4$ \\
\hline
\end{tabular}
\end{table*}

The survey selection function $W(\vec{x})$ expresses the expected mean
density of galaxies with spectroscopic redshifts at position
$\vec{x}$, given the angular and luminosity survey selection criteria.
An accurate determination of the selection function is essential in
order to estimate the power spectrum, which describes the amplitude of
fluctuations relative to the mean density.

Determination of the WiggleZ survey selection function is complicated
by several factors, which are discussed in the following sub-sections:
\begin{itemize}
\item The boundaries of each survey region are determined by the
  overlapping network of UV and optical imaging coverage [see Section
  \ref{seccover}].
\item The magnitude cuts used to select galaxies from the input
  imaging lie close to the faint completeness thresholds of those
  surveys.  For example, the completeness of the GALEX UV data at the
  survey flux threshold varies with the amount of Galactic foreground
  dust, inducing a variation in target density with angular position
  [see Section \ref{secparent}].
\item The spectroscopic redshift completeness of each AAOmega pointing
  (i.e.\ the fraction of observed spectra producing successful
  redshifts) varies considerably with observing conditions such as
  astronomical seeing and cloud cover, inducing a variation in the
  density of successful redshifts with angular position [see Section
  \ref{secredcomp}].
\item The program of GALEX imaging has proceeded simultaneously with
  the spectroscopic follow-up at the AAT.  The expansion of the
  angular mask of the survey with time must be tracked in order to
  properly model the angular density of redshifts [see Section
  \ref{secredcomp}].
\item The spectroscopic redshift completeness of each AAOmega pointing
  varies systematically across the spectrograph field-of-view,
  decreasing towards the edges where acquisition errors are amplified
  [see Section \ref{secfieldcomp}].
\item Whilst the survey is unfinished, the optical magnitude
  distribution of observed galaxies varies with position on the sky
  owing to the target prioritization scheme used for the observations
  (Drinkwater et al.\ 2010).  This implies that the galaxy redshift
  distribution also varies with angular position [see Section
  \ref{secnz}].
\item A fraction of the assigned galaxy redshifts are ``blunders''
  resulting from emission-line mis-identifications.  The rate of
  blunders depends on redshift, and hence distorts the true
  radial galaxy distribution [see Section \ref{secbadz}].
\end{itemize}

\subsection{Coverage masks}
\label{seccover}

{\it This part of the selection function establishes a (0,1) binary
  angular coverage mask indicating the availability of input targets.}

The boundaries of each WiggleZ survey region are defined by the
coverage maps of the UV and optical imaging data within that region.
The UV data consists of a series of pointings of the GALEX satellite.
The GALEX field-of-view is a circle of radius $0.6$ deg; we select
sources from a slightly smaller radius of $0.55$ deg due to concerns
about the GALEX photometry at the edges of the field (Morrissey et
al.\ 2007).  For the survey regions analyzed in this paper, the
optical coverage map corresponds to the 4th Data Release of the SDSS.

The GALEX source catalogues contain small regions of bad data
corresponding to scattered light from bright stars adjacent to the
field-of-view.  We inspect each GALEX field by eye for the presence of
these artefacts and define rectangular masks to remove this bad data.
These masks encompass a negligible fraction ($< 1\%$) of the survey
area.

\subsection{Variation of parent density with dust and exposure time}
\label{secparent}

{\it This part of the selection function modulates the angular
  coverage map in accordance with incompleteness in the parent imaging
  catalogues.}

The faint UV magnitude threshold for WiggleZ survey selection, $NUV =
22.8$, is comparable to the magnitude completeness limit of the GALEX
Medium Imaging Survey (MIS) which provides the targets.  The
incompleteness of the WiggleZ target catalogue is therefore
significant and must be modelled in our selection function, because it
determines the baseline target density on the sky in the absence of
clustering.  We calibrated this incompleteness factor as a function of
foreground Galactic dust (quantified by the value of $E_{B-V}$) and
GALEX exposure time $t_{\rm exp}$.

\begin{figure}
\center
\epsfig{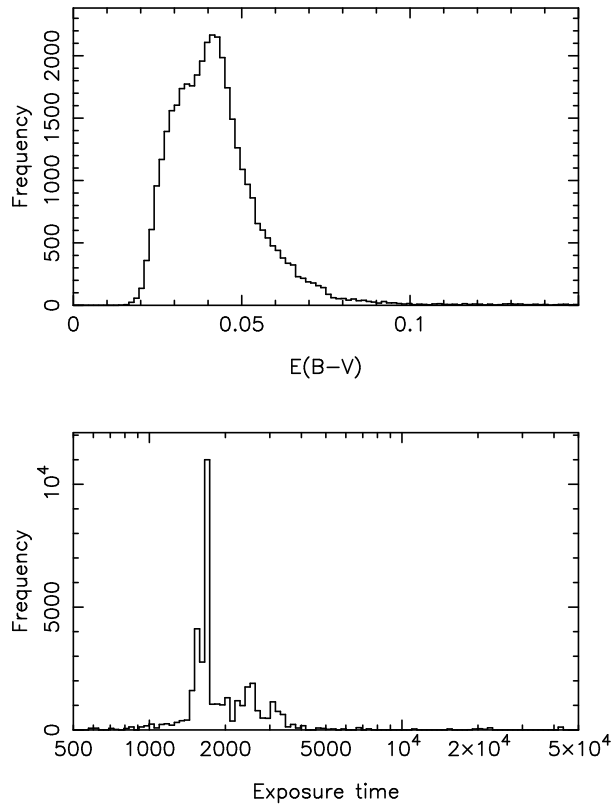}
\caption{Histograms of the values of GALEX exposure time and Galactic
  dust extinction across the WiggleZ survey regions analyzed in this
  paper.}
\label{fighist}
\end{figure}

The variations of $E_{B-V}$ and $t_{\rm exp}$ across the survey
regions analyzed in this paper are displayed in Figure \ref{fighist}.
The distribution of exposure time is peaked in the range $1400 <
t_{\rm exp} < 1800$ sec (the canonical MIS exposure time is 1500 sec,
equivalent to one orbit of the GALEX satellite when overheads are
included).  However, our analysis also includes fields with lower
exposure times $500 < t_{\rm exp} < 1400$ sec for which we are still
gathering data.  In addition some fields have significantly larger
values of $t_{\rm exp}$ if they have been observed as part of other
GALEX projects.  The distribution of dust extinction is broad, peaking
at $E_{B-V} \approx 0.04$.  Although we correct the UV magnitudes for
dust extinction (by an amount $A_{NUV} \approx 8 \, E_{B-V} \approx
0.3$, see Drinkwater et al.\ 2010), the proximity of our observations
to the MIS magnitude threshold induces a variation of completeness
with $E_{B-V}$.

We used the GALEX $NUV$ (dust-corrected) differential source counts
(i.e., counts of galaxies in $NUV$ magnitude bins) to calibrate the
dependence of the WiggleZ survey completeness on $E_{B-V}$ and $t_{\rm
  exp}$.  For this analysis we created a GALEX-SDSS matched galaxy
sample for which we did not impose any of the WiggleZ survey magnitude
or colour cuts.  This is because the WiggleZ selection cuts reduce the
target density by a factor of approximately 6, greatly increasing the
noise in these measurements.  We matched the GALEX NUV catalogues with
the SDSS data (using a tolerance of $2.5$ arcsec) and removed objects
flagged as stars in SDSS.  The presence of these stars would distort
the galaxy number counts and induce spurious correlations with
$E_{B-V}$ owing to the non-uniform stellar distribution.

We constructed the source count by dividing the survey coverage map
into small pixels of size $0.1 \times 0.1$ deg and assigning each
pixel a mean value of $E_{B-V}$ and $t_{\rm exp}$.  We then added up
the pixel source counts in bins of $E_{B-V}$ and $t_{\rm exp}$.

\begin{figure}
\center
\epsfig{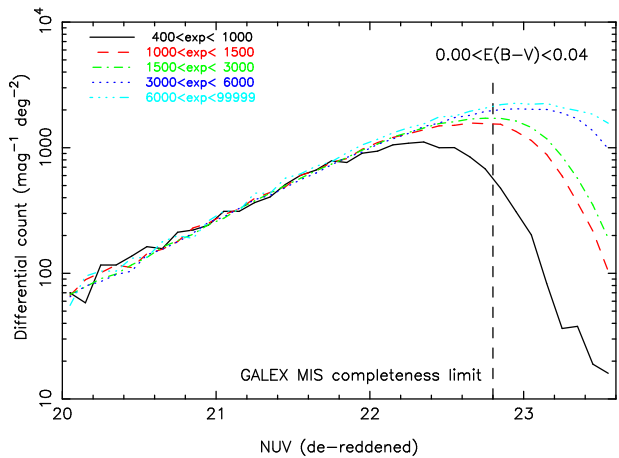}
\caption{Differential source counts of GALEX-SDSS galaxy matches in
  bins of GALEX exposure time.  We restrict the analysis to regions of
  the survey with low dust extinction $E_{B-V} < 0.04$.  We note that
  the uppermost curve with the caption ``$6000 < {\rm exp} < 99999$''
  corresponds to the ``fiducial'' source count in the limit of low
  dust extinction and high GALEX exposure time, as discussed in the
  text.}
\label{figdiffexp}
\end{figure}

Figure \ref{figdiffexp} illustrates the source count in bins of
exposure time, restricting the analysis to pixels with low dust
extinction ($E_{B-V} < 0.04$) in order to isolate the variation with
exposure time.  We expect to see the number counts rising with
increasing magnitude (in the classic Euclidean regime, $d({\rm
  log}_{10}N)/dm = 0.6$) and tailing off at faint magnitudes when
incompleteness becomes significant.  Indeed, Figure \ref{figdiffexp}
illustrates that the counts in different exposure time bins agree well
at bright magnitudes, and the completeness limit grows fainter with
increasing exposure time.  More detailed fits are presented below.

\begin{figure}
\center
\epsfig{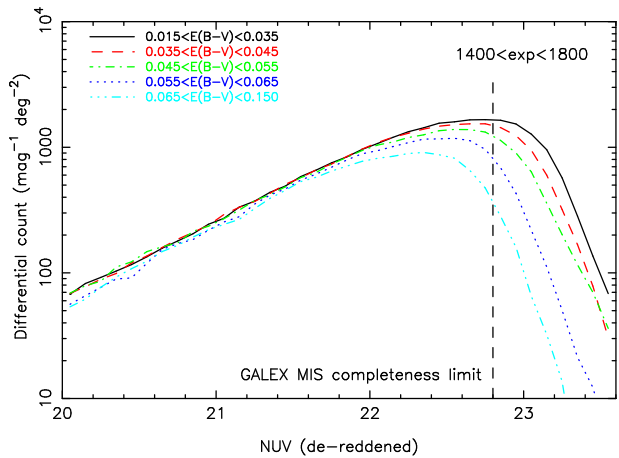}
\caption{Differential source counts of GALEX-SDSS galaxy matches in
  bins of dust extinction.  We restrict the analysis to regions of the
  survey with GALEX exposure times $1400 < t_{\rm exp} < 1800$ sec.}
\label{figdiffebv}
\end{figure}

Figure \ref{figdiffebv} displays the counts in bins of $E_{B-V}$ for
GALEX observations in the exposure time range $1400 < t_{\rm exp} <
1800$ sec, in order to isolate the variation with dust extinction.
Again, the agreement is good at bright magnitudes (i.e.\ there is no
significant residual dependence of the number counts on dust
extinction) and the incompleteness at faint magnitudes increases with
the value of $E_{B-V}$.

We found that a good fitting formula for these source counts as a
function of de-reddened magnitude $m = NUV$ is a power-law modulated
by an incompleteness function:
\begin{equation}
\frac{dN}{dm}(m) = \frac{dN_0}{dm}(m) \times C(\mu,\sigma,m)
\label{eqcounts1}
\end{equation}
where
\begin{eqnarray}
  \frac{dN_0}{dm}(m) &=& 10^{\alpha + m \beta} \label{eqpowerlaw} \\ C(\mu,\sigma,m) &=& 0.5 \left[ 1 + {\rm erf} \left( \frac{\mu - m}{\sigma} \right) \right] \label{eqcounts2}
\end{eqnarray}
where ${\rm erf}(x)$ is the error function:
\begin{equation}
{\rm erf}(x) = \frac{2}{\sqrt{\pi}} \int_0^x \exp{(-t^2)} \, dt
\end{equation}

We first fitted Equation \ref{eqcounts1} to the source counts for
survey regions in the ranges $E_{B-V} < 0.04$ and $t_{\rm exp} > 6000$
sec, which we designated the ``fiducial'' source count
$C_0(\mu_0,\sigma_0,m)$ in the limit of low dust extinction and high
GALEX exposure time.  The apparent incompleteness in this measurement
is not due to GALEX observations but to the limiting magnitude
threshold of the SDSS data with which we match the GALEX sources.  We
fitted the model to the magnitude range $20.0 < NUV < 22.8$ (motivated
by the faint flux threshold for WiggleZ target selection).  The
best-fitting parameters for the fiducial source count are $\beta =
0.625 \pm 0.013$, $\mu_0 = 22.98 \pm 0.05$ and $\sigma_0 = 1.1 \pm
0.1$, which we assumed for the rest of this analysis.  The errors in
each parameter are quoted after marginalizing over the remaining
parameters.  Our derived number counts are consistent with Xu et al.\
(2005), who show that a model incorporating luminosity evolution
provides a good fit to this data.  We note that the overall amplitude
of the number counts, parameterized by $\alpha$ in Equation
\ref{eqpowerlaw}, is not required for determining the incompleteness,
as explained below.

\begin{figure}
\center
\epsfig{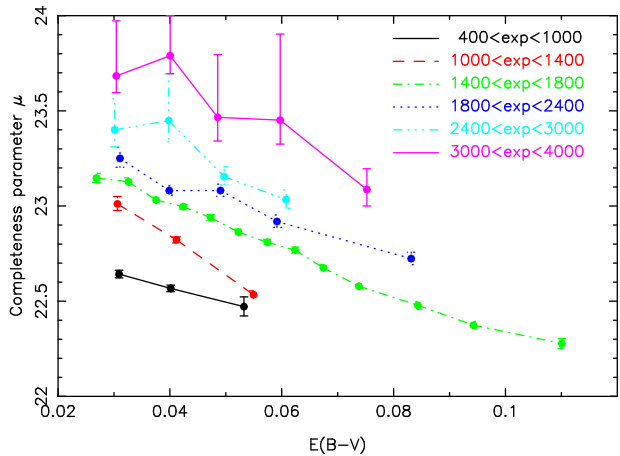}
\caption{Dependence of the completeness parameter $\mu$ on dust
  extinction, in bins of exposure time.  The measurements of $\mu$ are
  of highest quality for the dominant exposure time range $1400 <
  t_{\rm exp} < 1800$ sec.  The error range in $\mu$ is asymmetric
  about the best-fitting value because the model becomes increasingly
  insensitive to $\mu$ as the value of $\mu$ increases.  The
  measurement for each dust bin is plotted at the median value of
  $E_{B-V}$ for that bin.}
\label{figmufits}
\end{figure}

We then defined the incompleteness in the GALEX catalogues as a
function of $E_{B-V}$ and $t_{\rm exp}$, relative to these fiducial
counts, by fitting the model:
\begin{equation}
\frac{dN}{dm}(m) = \frac{dN_0}{dm}(m) \times C_0(\mu_0,\sigma_0,m)
\times C(\mu,\sigma,m)
\label{eqdndm}
\end{equation}
By construction, $C = 1$ in the limit of low dust extinction and high
exposure time.  We found that the best-fitting value of $\sigma$ does
not vary significantly with dust or exposure time.  Therefore we fixed
$\sigma = 0.578$ such that the completeness function depended on just
one parameter, $\mu$.  We then fitted the value of $\mu$ as a function
of both $E_{B-V}$ and $t_{\rm exp}$; the results are displayed in
Figure \ref{figmufits}, together with the error in $\mu$ obtained by
marginalizing over the other fitted parameters.

We can compare these results to theoretical expectations.  The
variation of the completeness limit $\mu$ with $t_{\rm exp}$ (at fixed
$E_{B-V}$) agrees well with the theoretical expectation for
background-limited imaging of fixed signal-to-noise, in which a
doubling of exposure time equates to approximately $0.4$ magnitudes of
survey depth.  The variation of $\mu$ with $E_{B-V}$ (at fixed $t_{\rm
  exp}$) should follow the dust-extinction law $\Delta \mu \approx 8
\Delta E_{B-V}$.  The observed slope in Figure \ref{figmufits} is in
fact a little steeper, owing to the influence of the additional factor
$C_0(m)$ in Equation \ref{eqdndm}.

The completeness of the GALEX-SDSS parent catalogue for the UV flux
threshold $m_0 = 22.8$ can be determined for a given $\mu(t_{\rm exp},
E_{B-V})$ by evaluating
\begin{equation}
{\rm Completeness}(\mu) = \frac{ \int_0^{m_0} \frac{dN_0}{dm} \,
  C_0(\mu_0,\sigma_0,m) \, C(\mu,\sigma,m) \, dm}{ \int_0^{m_0}
  \frac{dN_0}{dm} \, C_0(\mu_0,\sigma_0,m) \, dm}
\label{eqcomp}
\end{equation}

\begin{figure}
\center
\epsfig{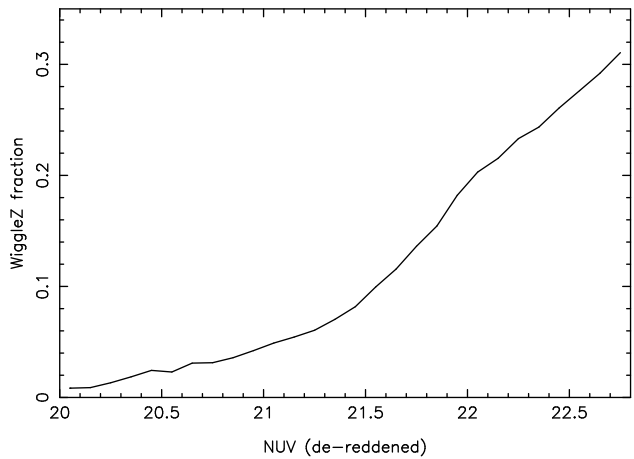}
\caption{Fraction of GALEX-SDSS matches that are selected as WiggleZ
  targets as a function of GALEX $NUV$ magnitude.}
\label{fignuvfrac}
\end{figure}

In order to model the WiggleZ survey target density from these
completeness measurements we must now allow for the fraction of
GALEX-SDSS matched sources that are selected as WiggleZ targets as a
function of $NUV$ magnitude.  This function is plotted in Figure
\ref{fignuvfrac}.  WiggleZ targets are preferentially faint in $NUV$
owing to the colour selection cut $-0.5 < NUV-r < 2$.  We weighted
Equation \ref{eqcomp} with this function in order to determine the
WiggleZ survey angular completeness map.  We defined this map using a
uniform grid of pixels in right ascension and declination.

\begin{figure}
\center
\epsfig{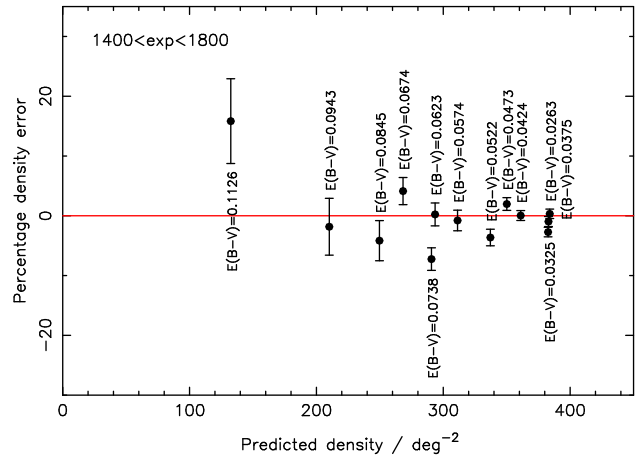}
\caption{Comparison of the observed and modelled target density for
  the WiggleZ catalogue in bins of dust extinction for the GALEX
  exposure time range $1400 < t_{\rm exp} < 1800$ sec.  The average
  value of $E_{B-V}$ for each dust bin is displayed next to the
  corresponding data point.  Poisson error bars, which are likely to
  be a mild under-estimate of the true error, are shown for the
  measured density.}
\label{figdenscomp}
\end{figure}

\begin{figure*}
\center
\epsfig{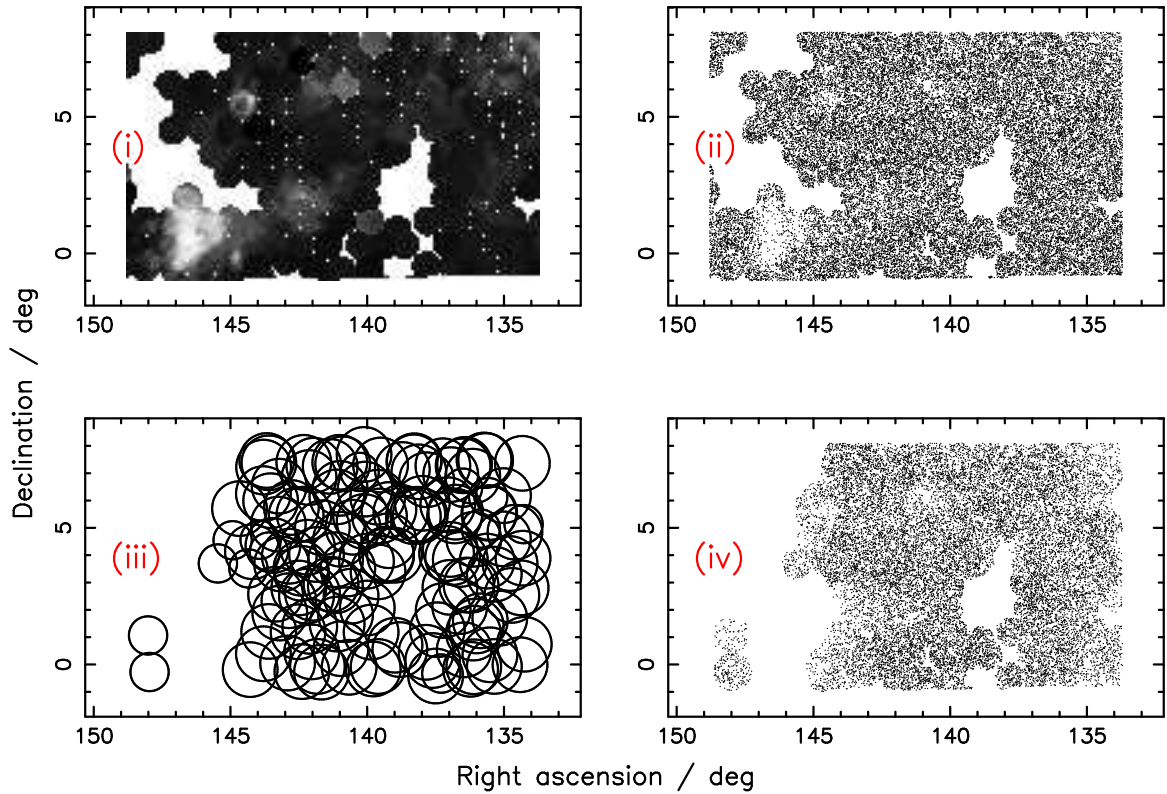}
\caption{Panels showing for the WiggleZ 9-hr region: (i) the angular
  completeness map of the parent catalogue, (ii) a realization of the
  parent catalogue sampled from this map, (iii) the sequence of 2dF
  field centres, (iv) a realization of the redshift catalogue.  Note
  in panel (iii) that in some pointings of the survey a field radius
  of $0.7$ deg rather than $1.0$ deg was used, due to concerns over
  radial-dependent redshift completeness within 2dF pointings.}
\label{figreal}
\end{figure*}

In Figure \ref{figdenscomp} we compare the WiggleZ target densities as
a function of dust extinction predicted by our number-counts modelling
with the densities observed in the catalogue.  We plot the percentage
difference of the measured density from the predicted density.  The
good match indicates that our selection function model is successful.
An example parent catalogue completeness map for the 9-hr survey
region is displayed as a panel in Figure \ref{figreal}.

We note that the variation in the angular density of targets is
dominated by incompleteness in the UV imaging data rather than in the
optical imaging data.  Although the faint optical magnitude threshold
for WiggleZ selection ($r = 22.5$) lies at the completeness limit of
the SDSS, the $NUV-r$ colour selection cut implies that the median
$r$-band magnitude of WiggleZ targets is $r \approx 21.5$ and the
variation of SDSS completeness with $E_{B-V}$ is negligible.  We also
studied the variation of completeness with the local astronomical
seeing in the SDSS images, and found no significant effect on the
target densities.

\subsection{Redshift completeness map}
\label{secredcomp}

{\it This part of the selection function samples the density map of
  the parent target catalogue with the pattern of spectroscopic
  follow-up observations.}

The spectroscopic observations of the WiggleZ survey are defined by a
series of AAT pointing centres across each survey region.  The
pointing centres for each region are determined prior to each survey
observing run based upon the available distribution of targets at the
time, using a ``simulated annealing'' algorithm.  Each patch of sky
must be observed three to four times on average to build up the
WiggleZ redshift catalogue, therefore we obtain a series of
overlapping pointings.  Each individual pointing results in a fraction
of successful redshifts which varies considerably with astronomical
seeing, airmass and cloud cover from under $40\%$ in poor conditions
($> 2.5$ arcsec seeing or airmass ${\rm sec}(z) > 1.5$ or $> 2/8$ths
cloud) to over $80\%$ in good conditions ($< 1.5$ arcsec seeing and
${\rm sec}(z) < 1.4$ and no cloud).  Furthermore, the fraction of
successful redshifts in any pointing varies across the field plate of
the 2dF spectrograph, as discussed in more detail below.  Therefore,
as the survey is still partially complete, the redshift completeness
map is a complicated function of position on the sky.

We determined the redshift completeness function in each survey region
using Monte Carlo realizations of the observations.  The steps in the
construction of a realization are illustrated in Figure \ref{figreal}.
For each region we started with the density map as a function of dust
and exposure time created by the process described in Section
\ref{secparent} [Figure \ref{figreal}, panel (i)].  We created a Monte
Carlo realization of this density map, containing the same number of
galaxies as the real target catalogue, by first generating a uniform
random catalogue and then excluding points based on a probability
equal to the local incompleteness [Figure \ref{figreal}, panel (ii)].
The spectroscopic observations consist of a series of overlapping AAT
pointings [Figure \ref{figreal}, panel (iii)].  Each pointing was
applied to our realization by laying down a field circle at the
correct angular position and randomly assigning the appropriate number
of observed sources to spectroscopic fibres.

We note that the available target area has grown over the duration of
the survey because the GALEX imaging observations are proceeding
simultaneously with the redshift follow-up.  In the Monte Carlo
simulations it is therefore necessary to track the distribution of
GALEX tiles available prior to each WiggleZ observing run and only
assign target sources to each random realization in areas that were
available at the time of each pointing.

For each telescope pointing in the random realization, a fraction of
the observed sources are flagged with successful redshifts.  The
remainder are flagged with unsuccessful redshifts.  The total number
of successful redshifts for each pointing in the Monte Carlo
realization is equal to that achieved in the corresponding survey
observation.  The probability of a given random source being assigned
a successful redshift varies across the field-of-view of the pointing
in the manner described below in Section \ref{secfieldcomp}.  The
resulting random realization, constructed by applying the full set of
pointings, is displayed in Figure \ref{figreal}, panel (iv).

The WiggleZ survey observing strategy involves obtaining repeat
spectra of unsuccessful redshifts observed in poor conditions.  In
addition, a small number of galaxies with existing successful
redshifts are re-observed in order to quantify the redshift blunder
rate.  Each of these categories of observation are included when
constructing the Monte Carlo realizations.

\subsection{Variation of redshift completeness across the spectrograph field-of-view}
\label{secfieldcomp}

\begin{figure*}
\center
\epsfig{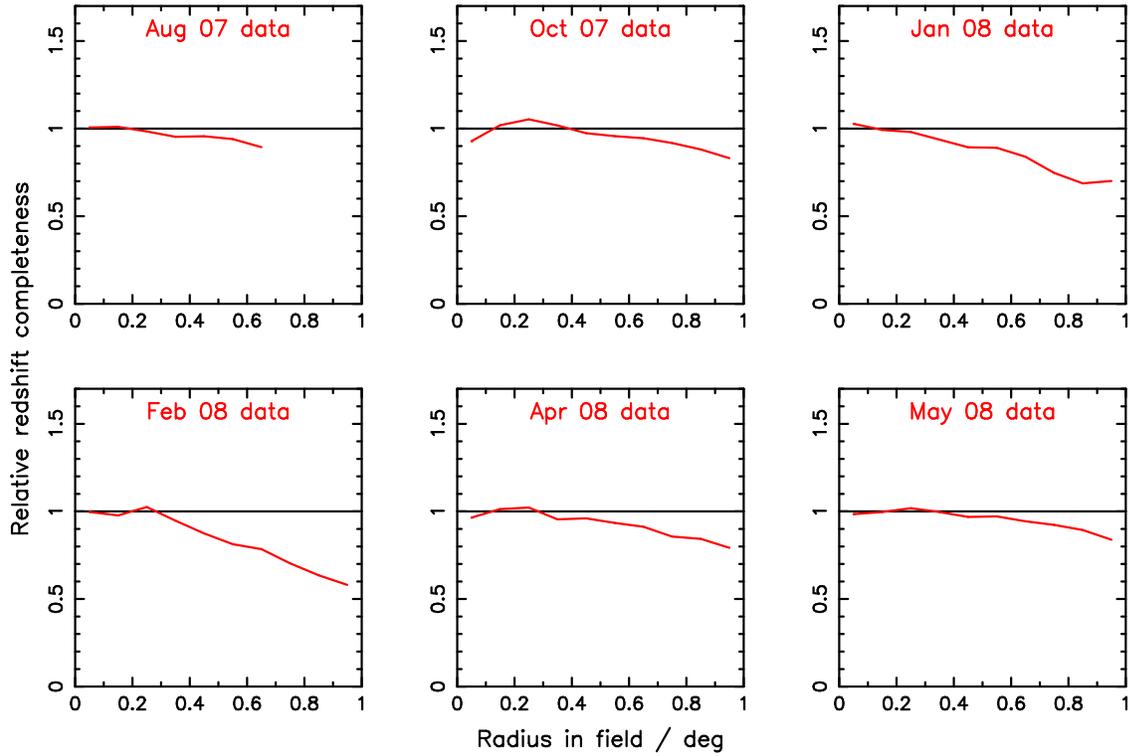}
\caption{Radial dependence of the redshift completeness in the
  2-degree field for the six WiggleZ survey observing runs between
  August 2007 and May 2008.  In August 2007 the field of observation
  was restricted to radii $< 0.7$ degrees.  We note that the
  completeness is normalized relative to a level of $1.0$ at the
  centre of each field.}
\label{figredcomprad}
\end{figure*}

\begin{figure*}
\center
\epsfig{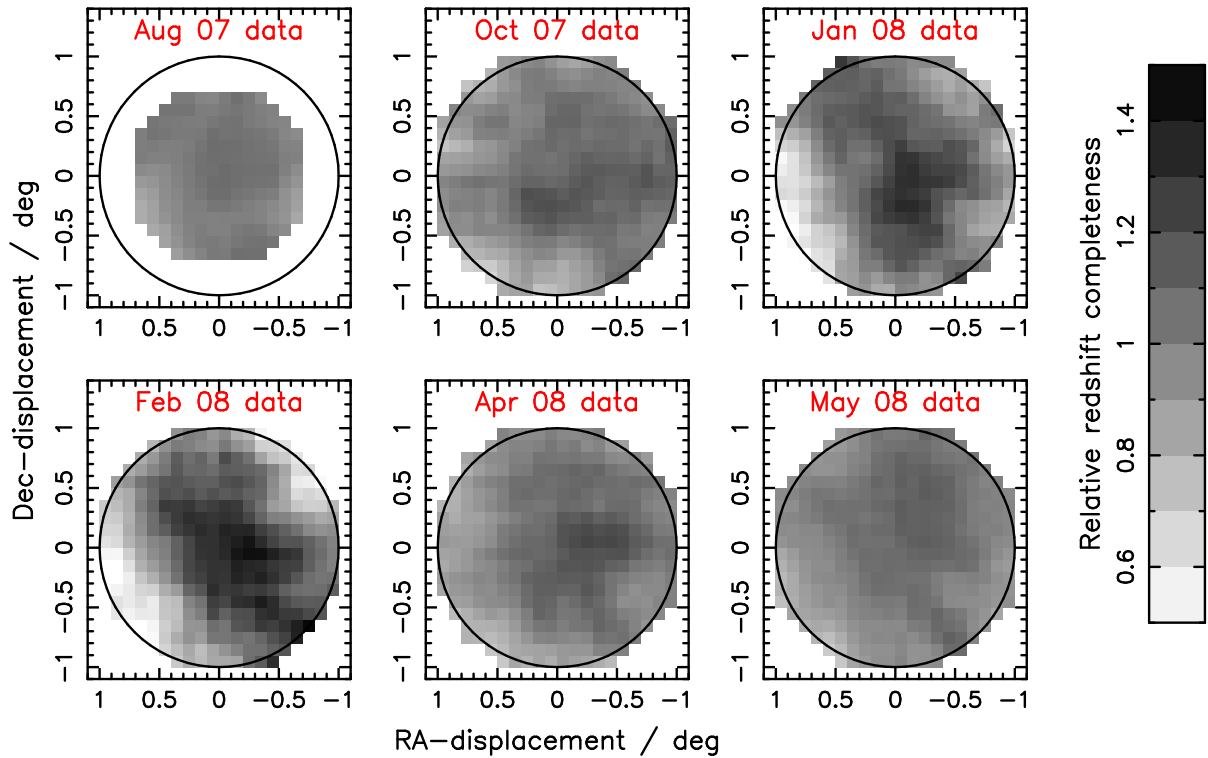}
\caption{Dependence of the redshift completeness on position in the
  2-degree field for the six WiggleZ survey observing runs between
  August 2007 and May 2008.  In August 2007 the field of observation
  was restricted to radii $< 0.7$ degrees.  We note that the
  completeness is normalized relative to a mean level of $1.0$
  averaged across each field.}
\label{figredcomp2d}
\end{figure*}

The probability of obtaining a successful redshift in a WiggleZ survey
pointing depends on the source position within the spectrograph
field-of-view.  We find that galaxies observed at the edges of the
field plate have a reduced redshift completeness because these
observations are more seriously affected by rotational mis-alignments
in target acquisition, which is performed using a limited number of
guide-fibre bundles.

Furthermore, we find that the pattern of redshift completeness across
the field-of-view varies between survey observing runs as the
spectrograph set-up is re-calibrated, but remains stable throughout an
observing run.  Although individual fibres can be placed in the
field-of-view by the 2dF robot positioner with an accuracy of 20$\mu$m
(0.3 arcsec), systematic errors can develop in the mapping of $(x,y)$
position on the field plate to (R.A., Dec.)  position on the sky.
These errors are caused by minor discrepancies in the atmospheric
correction applied to target apparent positions, and (at times during
the survey) by an instability in the modelling of opto-mechanical
distortions between the prime focus corrector and the 2dF field
plates.  These additional errors imply that the redshift completeness
variation is not radially symmetric.

These effects are illustrated by Figures \ref{figredcomprad} and
\ref{figredcomp2d}.  In Figure \ref{figredcomprad} we display the
radial variation in redshift completeness for WiggleZ survey observing
runs between August 2007 and May 2008.  Figure \ref{figredcomp2d}
displays the 2D variation of redshift completeness across the field
plate for each of these observing runs, and we use these 2D maps to
generate the probabilities of successful redshifts in the Monte Carlo
realizations described above.  In some observing runs, such as August
2007, we restricted our spectroscopic observations to the central
$0.7$ deg radius of the 1-deg radius field-of-view of the 2dF
spectrographs, because the drop in redshift completeness at the edges
of the field was particularly significant.

\subsection{Radial selection function versus angular position}
\label{secnz}

{\it This part of the selection function creates the appropriate
  radial distribution of galaxies depending on the distribution of
  $r$-band magnitudes in each 2dF pointing.}

The final step in the creation of the Monte Carlo realizations is to
assign a random redshift to each galaxy, thus establishing the
selection function in the radial direction.  Targets in the WiggleZ
survey are prioritized for observation in accordance with their SDSS
$r$-band magnitudes such that the faintest galaxies are observed
earliest in the sequence (Drinkwater et al.\ 2010).  In detail we use
five priority bands which divide the magnitude range $20.0 < r < 22.5$
into equal pieces.  Given that there is a correlation between redshift
and magnitude, this implies that the survey redshift distribution
varies across the sky in a manner dependent on the density of
redshifts obtained in a given area.

We track this in our random catalogues by recording for each telescope
pointing the number of successful redshifts obtained in each magnitude
priority band, and assigning these magnitude identifications to
sources in the random catalogues.  We note that the weak dependence of
redshift completeness on $r$-band magnitude is also absorbed into this
step of the process.

\begin{figure}
\center
\epsfig{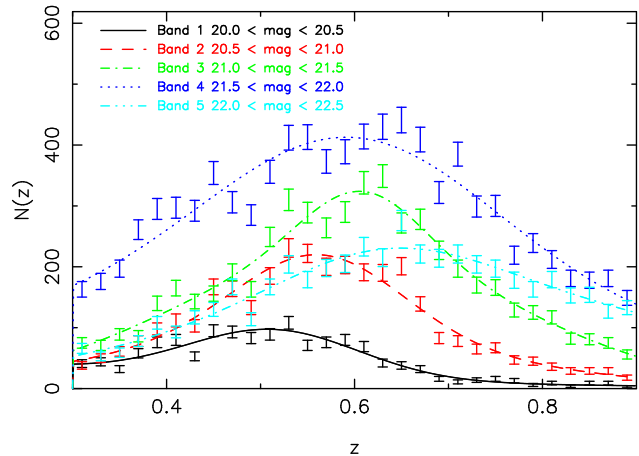}
\caption{The distribution of WiggleZ redshifts $z$ in the five
  magnitude bands in which the targets are prioritized.  Data is shown
  for the combined 9-hr, 11-hr and 15-hr survey regions for the
  redshift range $0.3 < z < 0.9$ analyzed in this paper.  Poisson
  error bars are shown for the points and the data is fitted with a
  sum of two Gaussian functions.}
\label{fignz}
\end{figure}

For each magnitude band we measured the redshift distribution $N(z)$
of successful redshifts using the existing WiggleZ spectroscopic data.
In order to reduce the fluctuations due to cosmic variance, we
combined the data for the 9-hr, 11-hr and 15-hr survey regions in this
analysis.  We used these probability distributions to assign redshifts
to random sources on the basis of their magnitude.  Figure \ref{fignz}
plots $N(z)$ for the five magnitude bands for the combined WiggleZ
regions.  We fit the redshift distributions with a sum of two Gaussian
functions and sample the random redshifts from these smooth
distributions.  Future work will establish $N(z)$ from measurements of
the galaxy luminosity function instead.

\subsection{Final construction of the survey selection function}

The method described in the preceding sub-sections allows us to
construct Monte Carlo realizations for each WiggleZ region
incorporating the angular and radial variations of the selection
function.  We can accurately determine the full selection function
function $W(\vec{x})$ by stacking together many such realizations: we
typically generate $10{,}000$ realizations for each region.  We
pixelize the selection function in grid cells corresponding to our
power spectrum measurement, as discussed in Section \ref{secfkp}.
When converting redshifts to distances we use a fiducial flat
$\Lambda$CDM cosmological model with matter density $\Omega_{\rm m} =
0.3$.

\subsection{Redshift blunder rate}
\label{secbadz}

The redshifts of galaxies in the WiggleZ survey are typically based on
identifications of emission lines; the signal-to-noise of the spectra
is usually too low to permit detection of the galaxy continuum.  The
principal line used for redshift identification is the [OII] doublet
at rest-frame wavelength 3727\AA.  This emission line lies in our
observed spectral window $4700 - 9500$\AA\ for the galaxy redshift
range $0.26 < z < 1.55$.  The redshift identification is confirmed for
most galaxies by the additional presence of emission lines such as
H$\beta$ 4861\AA, [OIII] 4959\AA, [OIII] 5007\AA, and H$\alpha$
6563\AA\ (Drinkwater et al.\ 2010).

However, not all redshift identifications are based on multiple
emission lines.  Features redward of [OII] progressively leave the
observed spectral range with increasing redshift.  The H$\beta$ and
H$\alpha$ lines are observable for the ranges $z < 0.95$ and $z <
0.45$, respectively.  At relatively high redshifts the galaxy emission
lines must be identified against a background of noisy sky emission
lines.  Despite these difficulties, we can gain some confidence in
single-line redshifts based on [OII] either through detection of the
doublet, which is marginally possible with our spectral resolution for
galaxies lying at $z > 0.8$, or by eliminating other solutions by
failure to detect [OII] at lower wavelengths in cleaner parts of the
spectrum.

We assign quality flags from $Q=1$ (lowest) to $Q=5$ (highest) for
each WiggleZ redshift based on the confidence of our measurement.
Redshifts with quality $Q \ge 3$ are considered ``reliable'' and used
in our analysis.  Redshifts with quality $Q \ge 4$ are based on
multiple emission lines and are very secure.  Galaxies with redshifts
based on noisy data or single emission lines are assigned $Q = 3$.
The fraction of reliable redshifts with $Q = 3$ is approximately
one-third.

Some fraction of WiggleZ redshifts will be blunders.  We distinguish
two types of redshift blunder for the purposes of our analysis.
Firstly, a galaxy emission line may be mis-identified as another,
incorrect, emission line.  In our power spectrum measurement this
represents (approximately) a convolution of the galaxy density field
whereby structures at a given redshift are coherently copied to a
second redshift.  Secondly, a night-sky emission line may be
mis-identified as a galaxy emission line.  As there are a large number
of night-sky emission lines available for mis-identification, this
effectively corresponds to a randomizing of the galaxy density field
through the addition of objects whose positions are uncorrelated with
the underlying density.

We studied the redshift blunder rate through a programme of repeat
observations.  In each survey pointing we assigned a small number of
spectrograph fibres (typically 3-5) to galaxies which have already
been assigned redshifts with quality $Q \ge 3$.  We define two
redshifts as inconsistent if they differ by $\Delta z > 0.002$ (the
typical redshift error for our spectra is $\Delta z = 0.0005$ or 100
km s$^{-1}$).

We find that pairs of repeat galaxy redshifts which both possess
quality $Q \ge 4$ disagree in $2\%$ of cases.  Assuming that one of
the pair of inconsistent values is the correct redshift, this implies
that the blunder rate for the set of $Q \ge 4$ redshifts is $1\%$.
Pairs of repeat redshifts which both possess $Q=3$ disagree in $31 \pm
2\%$ of cases.  However, we can obtain a larger statistical sample for
analysis if we consider pairs composed of $Q=3$ and $Q \ge 4$
redshifts, supposing that the higher quality redshift is the correct
value.  Under this method we find that the blunder rate of $Q=3$
redshifts is $17 \pm 1\%$, in good agreement with the internal blunder
rate for $Q=3$ pairs.  Given that approximately one-third of reliable
redshifts are assigned $Q=3$, the overall blunder rate for the WiggleZ
survey is about $5\%$.  However, we must carefully quantify the
redshift blunders in more detail in order to obtain an unbiased
measurement of the galaxy power spectrum.

\begin{figure}
\center
\epsfig{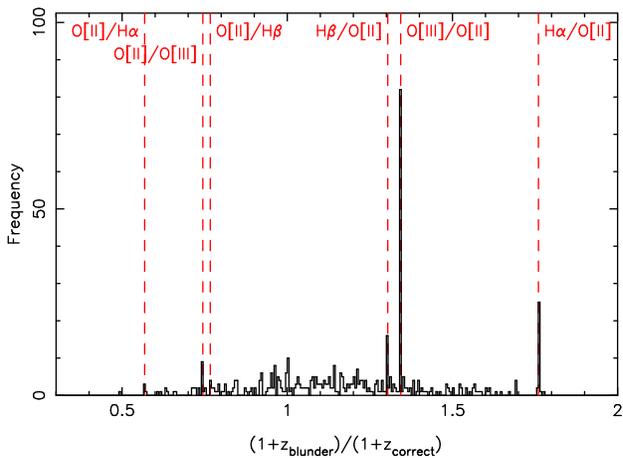}
\caption{Distribution of values of $(1+z_1)/(1+z_2)$ for inconsistent
  repeat redshifts derived from pairs of spectra with quality flags
  $Q=3$ (redshift $z_1$) and $Q \ge 4$ (redshift $z_2$).  The vertical
  lines indicate the ratios expected in the cases where H$\beta$,
  [OIII] and H$\alpha$ are mis-identified as [OII].}
\label{figzblundline}
\end{figure}

In Figure \ref{figzblundline} we illustrate the fraction of redshift
blunders resulting from the mis-identification of an emission line
with a second, incorrect, emission line by plotting the distribution
of values of $(1+z_1)/(1+z_2)$ for inconsistent repeat redshifts
composed of $Q=3$ and $Q \ge 4$ pairs.  The values of $(z_1,z_2)$ are
the redshifts of the $Q=3$ and $Q \ge 4$ spectra, respectively.  This
histogram reveals three significant spikes corresponding to the
mis-identification of H$\beta$, [OIII] 5007\AA\ and H$\alpha$ as
[OII].  Approximately $30\%$ of redshift blunders correspond to
this type of mis-identification; the correct redshift in such cases is
typically lower than the blunder redshift.

\begin{figure}
\center
\epsfig{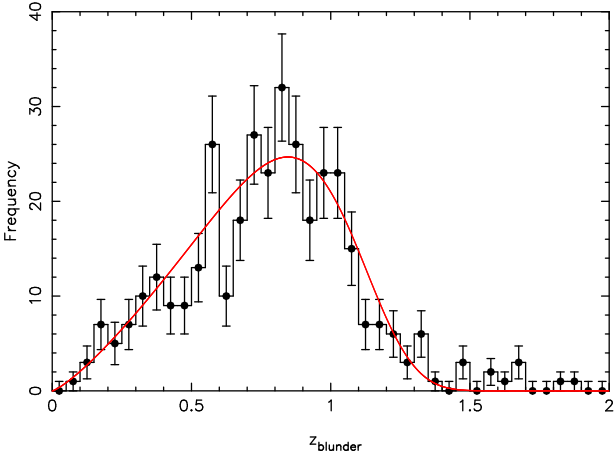}
\caption{The distribution of redshift blunders for the pairs in Figure
  \ref{figzblundline} which do not lie in the three prominent spikes.
  The solid line is the best fit of the model described by Equation
  \ref{eqnzfit}.}
\label{figzblundsky}
\end{figure}

Figure \ref{figzblundsky} plots the distribution of redshift blunders
not contained in the three spikes in Figure \ref{figzblundline}.  This
type of blunder, comprising about $70\%$ of all blunders, corresponds
to mis-identification of sky emission lines as [OII].  In Figure
\ref{figzblundsky} we have also fitted a model for the redshift
distribution $N(z)$ of the form:
\begin{equation}
N(z) \propto \left( \frac{z}{z_0} \right)^\alpha \exp{ \left[ - \left( \frac{z}{z_0} \right)^\beta \right] } .
\label{eqnzfit}
\end{equation}
The best-fitting parameters are $z_0 = 1.11$, $\alpha = 1.24$, $\beta
= 6.46$.  We note that the distribution of blunders peaks at a
significantly higher redshift than the distribution of correct
redshifts shown in Figure \ref{fignz}.  This implies that the blunder
fraction varies significantly with redshift -- this behaviour is
plotted in Blake et al.\ 2009, figure 6.  In Section \ref{secbadzcorr}
we model the effect of these types of redshift blunders on
measurements of the galaxy power spectrum.

\section{Power spectrum analysis}
\label{secpowerspec}

\subsection{Power spectrum estimation methodology}
\label{secfkp}

In this Section we summarize our method of power spectrum estimation,
prior to presenting our analysis of the WiggleZ survey data in Section
\ref{secpkmeas}.  Our power spectrum estimation is based on the
optimal weighting scheme of Feldman, Kaiser \& Peacock (1994) [FKP]
(also see the discussions in Tadros \& Efstathiou 1996; Hoyle et
al.\ 2002).  When converting redshifts to distances we use a fiducial
flat $\Lambda$CDM cosmological model with matter density $\Omega_{\rm
  m} = 0.3$.  We first enclosed the survey cone for the particular
region and redshift interval within a cuboid of sides $(L_x,L_y,L_z)$
and gridded the galaxy catalogue in cells numbering $(n_x,n_y,n_z)$
using nearest grid point (NGP) assignment to produce a distribution
$n(\vec{x})$.  The cell dimensions were chosen such that the Nyquist
frequencies in each direction (e.g.\ $k_{{\rm Nyq},x} = \pi n_x /
L_x$) exceeded the maximum frequency of measured power $k_{\rm max}$
by at least a factor of two (although we corrected for the effect of
NGP assignment on the power spectrum as explained below).

We then applied a Fast Fourier transform to the grid of data,
weighting each pixel by a factor $w(\vec{x})$ following the method of
FKP:
\begin{equation}
  \tilde{n}(\vec{k}) = \sum_{\vec{x}} n(\vec{x}) w(\vec{x}) \exp{ (i \vec{k}.\vec{x}) }
\end{equation}
where the weighting factor is given by
\begin{equation}
w(\vec{x}) = \frac{1}{1 + W(\vec{x}) N_c n_0 P_0}
\label{eqweight}
\end{equation}
(FKP, equation 2.3.4).  In Equation \ref{eqweight}, $N_c = n_x n_y
n_z$ is the total number of grid cells, $n_0 = N/V$ is the mean number
density of galaxies in units of $h^3$ Mpc$^{-3}$, where $N$ is the
total number of galaxies and $V = L_x L_y L_z$ is the cuboid volume,
and $P_0$ is a characteristic value of the power spectrum at the
Fourier wavescale of interest.  The purpose of the weighting scheme of
Equation \ref{eqweight} is give equal weight per volume where we are
limited by cosmic variance, and equal weight per galaxy where we are
limited by shot noise, resulting in an optimal measurement of the
power spectrum amplitude $P(k)$ in the case where $P(k) \approx P_0$.

In Equation \ref{eqweight}, $W(\vec{x})$ is proportional to the survey
selection function determined in Section \ref{secsel}, which describes
the number of galaxies expected in each cell $\vec{x}$ in the absence
of clustering.  We assume here the normalization
\begin{equation}
\sum_{\vec{x}} W(\vec{x}) = 1 .
\end{equation}
We note that summations over cells $\vec{x}$ may be related to the
equivalent integration over volume elements $d^3\vec{x}$ by
\begin{equation}
  \frac{1}{N_c} \sum_{\vec{x}} W(\vec{x}) = \frac{1}{V} \int W(\vec{x}) \, d^3\vec{x} .
\end{equation}
We also apply a Fast Fourier transform to the survey selection function:
\begin{equation}
\tilde{W}(\vec{k}) = \sum_{\vec{x}} W(\vec{x}) w(\vec{x}) \exp{ (i \vec{k}.\vec{x}) }
\end{equation}
The estimate of the galaxy power spectrum for wavescale $\vec{k}$ is
then
\begin{equation}
P_{\rm est}(\vec{k}) = \frac{|\tilde{n}(\vec{k}) - N \tilde{W}(\vec{k})|^2 - 
N \sum_{\vec{x}} W(\vec{x}) w^2(\vec{x})}{N^2 N_c \sum_{\vec{x}} W^2(\vec{x}) w^2(\vec{x})}
\label{eqpkest}
\end{equation}
This estimate of the power spectrum is biased by the process of NGP
assignment (Jing 2005).  In order to remove this bias we calculated
the correction factor for each Fourier mode (by which the power
spectrum estimate should be divided):
\begin{equation}
{\rm Correction \; factor}(\vec{k}) = \frac{\sum_{\vec{m}}
  H^2(\vec{k'}) P(\vec{k'})}{P(\vec{k})}
\label{eqpkngpcorr}
\end{equation}
where $\vec{m} = (m_x,m_y,m_z)$ is a vector of integers, $\vec{k'} =
(k_x',k_y',k_z') = (k_x + m_x k_{{\rm Nyq},x}, k_y + m_y k_{{\rm
    Nyq},y}, k_z + m_z k_{{\rm Nyq},z})$ and
\begin{equation}
H(\vec{k}) = \frac{ \sin{ \left( \frac{\pi k_x}{2 k_{{\rm Nyq},x}}
    \right) } \sin{ \left( \frac{\pi k_y}{2 k_{{\rm Nyq},y}} \right) }
    \sin{ \left( \frac{\pi k_z}{2 k_{{\rm Nyq},z}} \right) }} { \left(
      \frac{\pi k_x}{2 k_{{\rm Nyq},x}} \right) \left( \frac{\pi
        k_y}{2 k_{{\rm Nyq},y}} \right) \left( \frac{\pi k_z}{2
        k_{{\rm Nyq},z}} \right) }
\end{equation}
In Equation \ref{eqpkngpcorr}, $P(\vec{k})$ is the underlying model
power spectrum.  Given that this is initially unknown, we proceed by
an iterative approach: we assume a fiducial cosmological model,
compute the correction factor, fit cosmological parameters to the
power spectrum, re-calculate the correction factor, and then repeat
the parameter fit.  The magnitude of the correction is typically $2\%$
at scale $k \approx 0.2 \, h$ Mpc$^{-1}$.

The spatially-varying selection function $W(\vec{x})$ has two effects
on the process of power spectrum estimation.  Firstly the expectation
value of Equation \ref{eqpkest} is the underlying power spectrum
$P(\vec{k})$ convolved with the survey selection function:
\begin{equation}
<P_{\rm est}(\vec{k})> = \frac{ \sum_{\vec{k'}} P(\vec{k'}) |\tilde{W}(\vec{k}-\vec{k'})|^2 }{ N_c \sum_{\vec{x}} W^2(\vec{x}) w^2(\vec{x}) }
\label{eqpkconv}
\end{equation}
The numerator of Equation \ref{eqpkconv} is summed over the grid
points $\vec{k'}$ in Fourier space for which the Fast Fourier
Transform of $W(\vec{x})$ is calculated, i.e.\ spaced by $(\Delta k_x,
\Delta k_y, \Delta k_z) = (2\pi/L_x, 2\pi/L_y, 2\pi/L_z)$.  For
reasons of computing speed when fitting models, we re-cast this
equation as a matrix multiplication in Fourier bins of width $\Delta k
= 0.01 h$ Mpc$^{-1}$:
\begin{equation}
<P_{\rm est}(i)> = \sum_j M_{ij} \, P_{\rm mod}(j)
\end{equation}
We determine the convolution matrix $M_{ij}$ by evaluating the full
sum of equation \ref{eqpkconv} for a set of unit vectors, e.g.\ for
bin $i$:
\begin{eqnarray}
P(\vec{k}) &=& 1 \; \; \; \; \; (k_{i,{\rm min}} < |\vec{k}| <
k_{i,{\rm max}}) \\ &=& 0 \; \; \; \; \; {\rm otherwise}
\end{eqnarray}

Secondly, the estimates of the power in different Fourier modes
$\vec{k}$ become correlated.  If we average the estimates of Equation
\ref{eqpkest} into bins in Fourier space, labelling the bins by $i$,
the covariance between bins $i$ and $j$ is given by
\begin{equation}
  < \delta P_i \, \delta P_j > = \frac{2}{N_{\vec{k}} N_{\vec{k'}}}
  \sum_{\vec{k},\vec{k'}} | P Q(\vec{k}-\vec{k'}) +
  S(\vec{k}-\vec{k'})|^2
\label{eqpkcov}
\end{equation}
(FKP, equation 2.5.2) where $\vec{k}$ and $\vec{k'}$ are pairs of
Fourier modes lying in bins $i$ and $j$, $P$ is the characteristic
power spectrum amplitude in bins $i$ and $j$ defined below, and the
functions $Q(\vec{k})$ and $S(\vec{k})$ are given by FKP equations
2.2.3 and 2.2.5:
\begin{eqnarray}
Q(\vec{k}) &=& \frac{ \sum_{\vec{x}} W^2(\vec{x}) w^2(\vec{x}) \exp{ (i \vec{k}.\vec{x}) }}{ \sum_{\vec{x}} W^2(\vec{x}) w^2(\vec{x}) } \\
S(\vec{k}) &=& \left( \frac{1}{n_0 N_c} \right) \frac{ \sum_{\vec{x}} W(\vec{x}) w^2(\vec{x}) \exp{ (i \vec{k}.\vec{x}) }}{ \sum_{\vec{x}} W^2(\vec{x}) w^2(\vec{x}) }
\end{eqnarray}
In deriving Equation \ref{eqpkcov} it is assumed that the power
spectrum factor $P$ which appears is effectively constant over Fourier
separations $\vec{k}-\vec{k'}$ which produce correlated estimates.
For our datasets the Fourier transform of the selection function,
$\tilde{W}(\vec{k})$, is sufficiently compact around $k=0$ that this
is a valid approximation.  We evaluated Equation \ref{eqpkcov} for
each survey region by a direct summation over Fourier modes in the FFT
grid.  Equation \ref{eqpkcov} depends on the underlying power
spectrum, which is initially unknown, in the same manner as equation
\ref{eqpkngpcorr}.  We again used an iterative approach whereby we
initially used a default model power spectrum to make this
calculation, and then replaced it using the fitted parameters.

In order to facilitate comparison with other studies it is useful to
take the limit of these equations in the case where the selection
function is constant, i.e.\ $W(\vec{x}) = 1/N_c$.  The power spectrum
estimator of Equation \ref{eqpkest} becomes
\begin{equation}
P_{\rm est}(\vec{k}) = \frac{|\tilde{n}(\vec{k}) - N \tilde{W}(\vec{k})|^2 - 
N}{N^2}
\end{equation}
and the covariance matrix in Equation \ref{eqpkcov} reduces to a
diagonal matrix with entries:
\begin{equation}
< (\delta P_i)^2 > = \frac{2}{N_{\vec{k}}} \left( P + \frac{1}{n_0} \right)^2
\label{eqpksimperr}
\end{equation}
where $N_{\vec{k}}$ is the number of Fourier modes lying in bin $i$.
Equation \ref{eqpksimperr} clarifies that there are two sources of
error in an estimate of the power spectrum: cosmic variance and shot
noise, represented by the two terms inside the bracket.

\subsection{Redshift blunder correction}
\label{secbadzcorr}

In this Section we calculate the distortion in the galaxy power
spectrum created by the types of redshift blunder described in Section
\ref{secbadz}.  In order to gain intuition we begin with a simple
model using the ``flat-sky approximation'' which supposes that
galaxies are scattered in position along a single axis of the cuboid
(which we take as the $x$-axis).  Defining $\delta(\vec{x})$ as the
galaxy overdensity in the cell at position $\vec{x} = (x,y,z)$, the
galaxy number distribution is given by
\begin{equation}
n(\vec{x}) = N W(\vec{x}) [ 1 + \delta(\vec{x}) ]
\label{eqnx}
\end{equation}
where $W(\vec{x})$ is the selection function normalized as above.  We
now suppose that a fraction $f$ of galaxies are scattered in position
along the $x$-axis such that their final $x$-position is drawn from a
probability distribution $V(x)$ (i.e., as described by Equation
\ref{eqnzfit} for our data).  This process creates a scattered galaxy
number distribution given by
\begin{equation}
S(\vec{x}) = N_1 V(x) \sum_{x'} n(x',y,z)
\label{eqsx}
\end{equation}
where the normalization constant $N_1$ can be calculated by requiring
that $\sum_{\vec{x}} S(\vec{x}) = f N$.  Equations \ref{eqnx} and
\ref{eqsx} have Fourier transforms
\begin{eqnarray}
  \tilde{n}(\vec{k}) &=& N \left[ \tilde{W}(\vec{k}) + \tilde{C}(\vec{k}) \right] \\
  \tilde{S}(\vec{k}) &=& f N \tilde{V}(k_x) \left[ \tilde{W}(0,k_y,k_z) + \tilde{C}(0,k_y,k_z) \right]
\end{eqnarray}
where we have defined the convolved density field
\begin{equation}
\tilde{C}(\vec{k}) = \sum_{\vec{k'}} \tilde{W}(\vec{k'}-\vec{k}) \, \tilde{\delta}(\vec{k'})
\end{equation}
It is convenient to subtract the selection function contribution:
\begin{eqnarray}
  \tilde{n}'(\vec{k}) &=& \tilde{n}(\vec{k}) - N \tilde{W}(\vec{k}) \\
  \tilde{S}'(\vec{k}) &=& \tilde{S}(\vec{k}) - f N \tilde{W}(0,k_y,k_z) \tilde{V}(k_x)
\end{eqnarray}
We now construct the combined density field $m(\vec{x}) = (1-f)
n(\vec{x}) + S(\vec{x})$ and write (for the purposes of this
calculation) a simple power spectrum estimator
\begin{equation}
  P_{\rm est}(\vec{k}) = \frac{|\tilde{m}'(\vec{k})|^2 - N}{N^2}
\label{eqpkestsimp}
\end{equation}
where $\tilde{m}'(\vec{k}) = (1-f) \tilde{n}'(\vec{k}) +
\tilde{S}'(\vec{k})$.  The terms needed to determine
$|\tilde{m}'(k)|^2$ are:
\begin{eqnarray}
|\tilde{n}'(\vec{k})|^2 &=& (1-f)^2 N^2 |\tilde{C}(\vec{k})|^2 \\
\tilde{n}'^*(\vec{k}) \tilde{S}'(\vec{k}) &=& f(1-f) N^2 \tilde{V}(k_x) P'(\vec{k}) \\
|\tilde{S}'(\vec{k})|^2 &=& f^2 N^2 |\tilde{C}(0,k_y,k_z)|^2 |\tilde{V}(k_x)|^2
\end{eqnarray}
where
\begin{equation}
P'(\vec{k}) = \sum_{\vec{k'}} P(\vec{k'}) \tilde{W}_{\vec{k'}-\vec{k}} \tilde{W}^*_{\vec{k'}-(0,k_y,k_z)}
\end{equation}
Writing the convolved power spectrum as $P_c(\vec{k}) =
|\tilde{C}(\vec{k})|^2$, the final result for the power spectrum
estimate is:
\begin{eqnarray}
P_{\rm est}(\vec{k}) &=& (1-f)^2 P_c(\vec{k}) + 2f(1-f) {\rm Re}[\tilde{V}(k_x) P'(\vec{k})] \nonumber \\ &+& f^2 P_c(0,k_y,k_z) |\tilde{V}(k_x)|^2
\label{eqpkblund1}
\end{eqnarray}
In the special case that the selection function and scattering functions
are constant we find that:
\begin{eqnarray}
P_{\rm est}(\vec{k}) &=& P(\vec{k}) \hspace{1cm} (k_x = 0) \\
&=& (1-f)^2 P(\vec{k}) \hspace{1cm} (k_x \ne 0)
\label{eqpkblundsimp}
\end{eqnarray}
If the Fourier modes $\vec{k}$ are binned in spherical shells as a
function of $k = |\vec{k}|$ then, as the value of $k$ increases, the
contribution of the mode with $k_x = 0$ becomes less important and
$P_{\rm est}(k) = (1-f)^2 P(k)$ is an increasingly good approximation.

We now consider the case where a fraction $f$ of redshifts are
scattered by a fixed displacement along the $x$-axis, i.e.\ a
convolution of the density field by a function $U$.  Equation
\ref{eqsx} is replaced by the relation
\begin{equation}
S(\vec{x}) = N_1 \sum_{x'} n(x',y,z) U(x-x')
\end{equation}
with the Fourier transform
\begin{equation}
  \tilde{S}(\vec{k}) = f N \tilde{U}(k_x) [ \tilde{W}(\vec{k}) + \tilde{C}(\vec{k}) ]
\end{equation}
Defining $\tilde{S}'(\vec{k}) = \tilde{S}(\vec{k}) - f N \tilde{U}(k_x)
\tilde{W}(\vec{k})$ we find that
\begin{eqnarray}
  P_{\rm est}(\vec{k})/P_c(\vec{k}) &=& (1-f)^2 + 2f(1-f) {\rm Re}[\tilde{U}(k_x)] \nonumber \\ &+& f^2 |\tilde{U}(k_x)|^2
\label{eqpkblund2}
\end{eqnarray}
For a simple shift $U(x) = \delta(x-x_0)$ we find that
\begin{equation}
P_{\rm est}(\vec{k}) = \left[ 1 - 2f(1-f) [1-\cos{(k_x x_0)} \right] P(\vec{k})
\end{equation}

These formulae are only approximations in the case of real data.
Firstly, the sky is curved and the redshift scatters do not happen
along a single axis of the cube.  Secondly, the blunder fraction
depends on redshift.  Thirdly, the blunders due to line confusion are
not a strict convolution of the density field, but a transformation in
redshift $z_1 \rightarrow z_2$ of the form $z_2 = C (1+z_1) - 1$,
where $C$ is a constant depending on the rest wavelengths of the
lines.  Fourthly, if a convolved redshift is shifted beyond the edge
of the density cube, it does not ``wrap around'' as required by
periodic boundary conditions.  Fifthly, FKP estimation of the power
spectrum is used rather than Equation \ref{eqpkestsimp}.

In order to measure the distortion of our measured power spectrum due
to redshift blunders for the real data, we therefore created Monte
Carlo simulations of galaxy catalogues with a known input power
spectrum and the same selection function $W(\vec{x})$ as each of our
survey regions.  We then applied the redshift blunder distributions of
Figures \ref{figzblundline} and \ref{figzblundsky} to the mock
catalogues and re-measured the power spectrum.  Comparison of the
input and output power spectra, averaging over many Monte Carlo
realizations, provided the correction factor due to redshift blunders.
We tested our code by reproducing the relations given in Equations
\ref{eqpkblund1} and \ref{eqpkblund2} in the flat-sky case.

\begin{figure}
\center
\epsfig{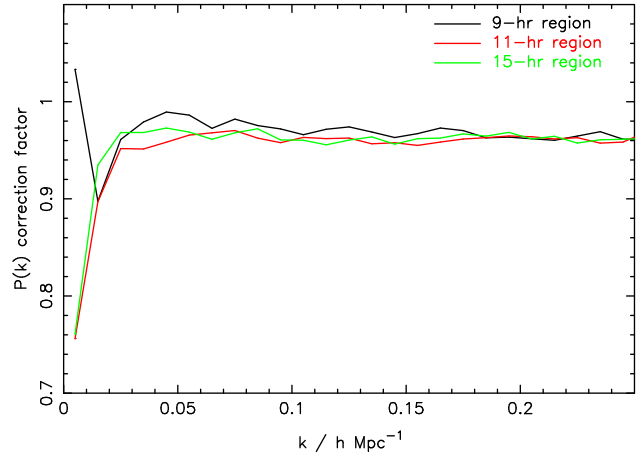}
\caption{The power spectrum correction factor due to redshift blunders
  for each of the survey regions analyzed in this paper, for a
  redshift range $0.3 < z < 0.9$.  The measured power spectrum must be
  divided by this factor in order to obtain an unbiased estimate of
  the true power spectrum.}
\label{figpkbadzcorr}
\end{figure}

Figure \ref{figpkbadzcorr} plots the correction factors for the
``angle-averaged'' power spectrum $P(k)$ for each of the survey
regions (taking a redshift interval $0.3 < z < 0.9$).  We see that a
good approximation for the correction for scales $k > 0.05 \, h$
Mpc$^{-1}$ is a constant, although a more significant correction is
required for large scales $k < 0.05 \, h$ Mpc$^{-1}$.  For $k > 0.05
\, h$ Mpc$^{-1}$ the approximately constant correction factor is not
exactly equal to $(1-f)^{-2}$ as predicted by Equation
\ref{eqpkblundsimp}, where $f$ is the average redshift blunder rate of
the catalogue; this is due to the mis-estimation of the denominator of
Equation \ref{eqpkest} that occurs because $W(\vec{x})$ is determined
from the galaxy redshift distribution including blunders, as described
in Section \ref{secnz}.  The Monte Carlo simulations performed here
also correct for this small bias in power spectrum estimation.

\subsection{Power spectrum measurement}
\label{secpkmeas}

\begin{figure*}
\center
\epsfig{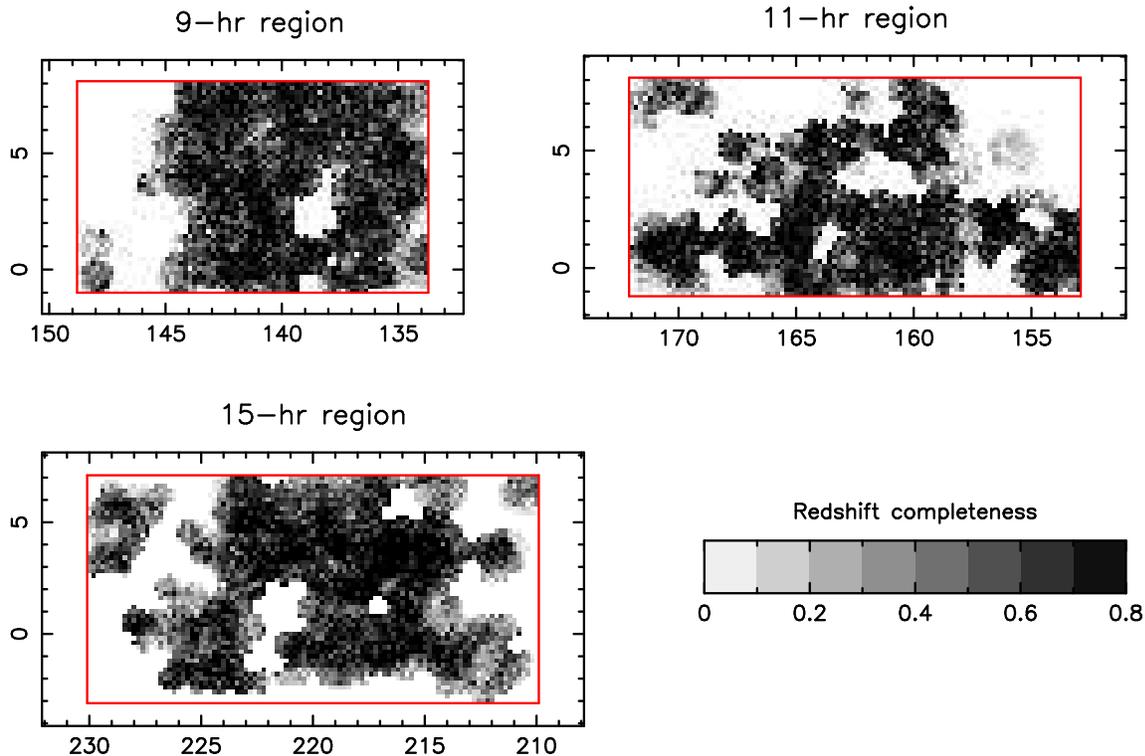}
\caption{Greyscale map illustrating the completeness of the
  spectroscopic follow-up of the WiggleZ targets for the three survey
  regions analyzed in this paper.  This Figure is generated by taking
  the ratio of the galaxy densities in the redshift and parent
  catalogues in small cells.  The $x$-axis and $y$-axis of each panel
  are right ascension and declination, respectively.}
\label{figradec}
\end{figure*}

In this study we analyzed a galaxy sample drawn from WiggleZ survey
observations prior to July 2009 in SDSS regions of our optical imaging
(9-hr, 11-hr, 15-hr).  Figure \ref{figradec} plots the (R.A., Dec.)
distribution of these redshifts.  We imposed the redshift cut $0.3 < z
< 0.9$ in order to remove the tails of the redshift distribution which
contain relatively few galaxies.  A total of $N = 56{,}159$ galaxy
redshifts remained.  We then split the sample into three redshift
slices $0.3 < z < 0.5$, $0.5 < z < 0.7$ and $0.7 < z < 0.9$.

We determined the effective redshift $z_{\rm eff}$ of our power
spectrum estimate in each redshift slice by weighting each pixel in
our 3D selection function by its contribution to the power spectrum
error:
\begin{equation}
z_{\rm eff}(k) = \sum_{\vec x} z \times \left( \frac{ n_g(\vec{x})
  P(k) }{ 1 + n_g(\vec{x}) P(k) } \right)^2
\label{eqzeff}
\end{equation}
where $n_g(\vec{x}) = (N_c N / V) W(\vec{x})$ is the galaxy number
density in each grid cell and $P(k)$ is the power spectrum amplitude.
In each case we used the best-fitting model power spectrum determined
below.  We evaluated this function at $k = 0.15 \, h$ Mpc$^{-1}$,
although the dependence on scale is weak.  The effective redshifts of
each slice determined using equation \ref{eqzeff} are $z_{\rm eff} =
(0.42, 0.59, 0.78)$.

We analyzed the three WiggleZ survey regions independently, resulting
in a total of nine power spectrum measurements.  We estimated the
power spectrum up to a maximum Fourier wavescale $k_{\rm max} = 0.4 \,
h$ Mpc$^{-1}$, assuming the value $P_0 = 2500 \, h^{-3}$ Mpc$^3$ for
the weighting factor in Equation \ref{eqweight}.  This choice is
motivated by our final measurement of the power spectrum amplitude
presented below on scales $k \approx 0.15 \, h$ Mpc$^{-1}$, but does
not have a strong influence on our results given that with the survey
partially complete the measurements are limited by shot noise on most
scales.  Representative values for the other parameters in Section
\ref{secfkp} are $(L_x,L_y,L_z) = (600,600,300) \, h^{-1}$ Mpc,
$(n_x,n_y,n_z) = (256,256,128)$, $V = 0.1 \, h^{-3}$ Gpc$^3$ and $n_0
= 5 \times 10^{-5} \, h^3$ Mpc$^{-3}$.  We combined the Fourier
amplitudes in angle-averaged bins of width $\Delta k = 0.01 \, h$
Mpc$^{-1}$.

\begin{figure*}
\center
\epsfig{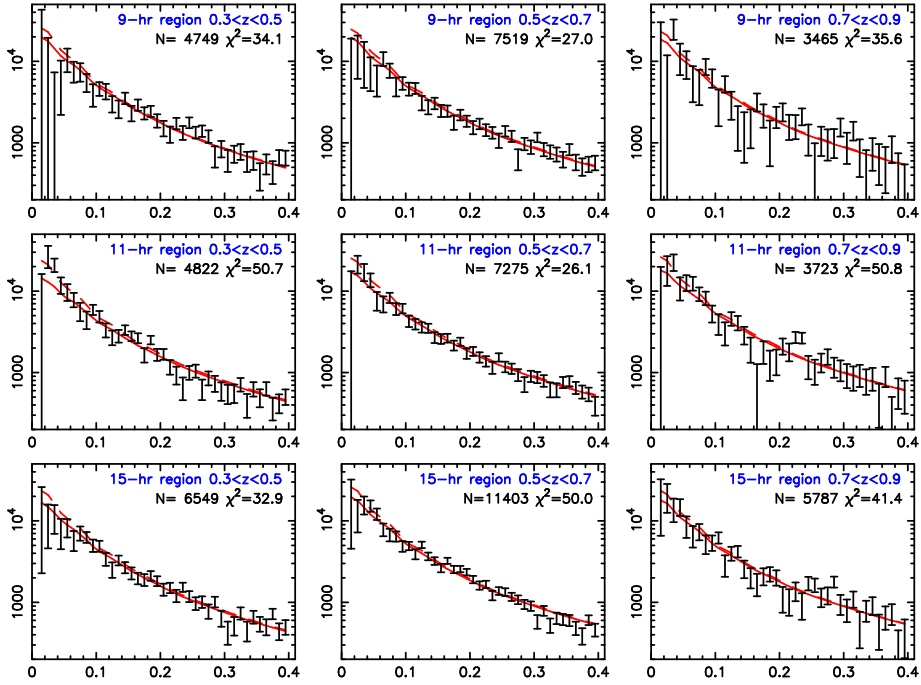}
\caption{The 1D galaxy power spectrum for each of the survey regions
  and redshift slices analyzed in this paper.  The data points in each
  plot are the power spectrum measured by FKP estimation.  The dashed
  line is an (unconvolved) model power spectrum evaluated by
  integrating equation \ref{eqpkmod} over angles.  We use cosmological
  parameters $\Omega_{\rm m} = 0.3$, $\Omega_{\rm b}/\Omega_{\rm m} =
  0.166$, $h = 0.72$, $n_{\rm s} = 0.96$ and $\sigma_8 = 0.8$,
  together with values of the redshift-space distortion parameters and
  linear bias factor fit to the 2D power spectrum for each redshift
  slice.  The solid line is the result of convolving this model power
  spectrum with the selection function for each region.  The
  chi-squared statistic is calculated over the range $k < 0.4 \, h$
  Mpc$^{-1}$ using the full covariance matrix.  The $x$-axis and
  $y$-axis of each panel are Fourier wavescale $k$ in units of $h$
  Mpc$^{-1}$ and power spectrum amplitude $P(k)$ in units of $h^{-3}$
  Mpc$^3$, respectively.}
\label{figpkreg}
\end{figure*}

The nine power spectrum measurements are plotted in Figure
\ref{figpkreg} together with a power spectrum model derived using a
``standard'' set of cosmological parameters together with a
prescription for redshift-space distortions.  The details of this
model are described below in Section \ref{secpkmod}.  The dashed and
solid lines illustrate the input model, and the model convolved with
the selection function for each region, respectively.  The model
provides an acceptable statistical fit to the measured power spectrum
in each case.

\begin{figure*}
\center
\epsfig{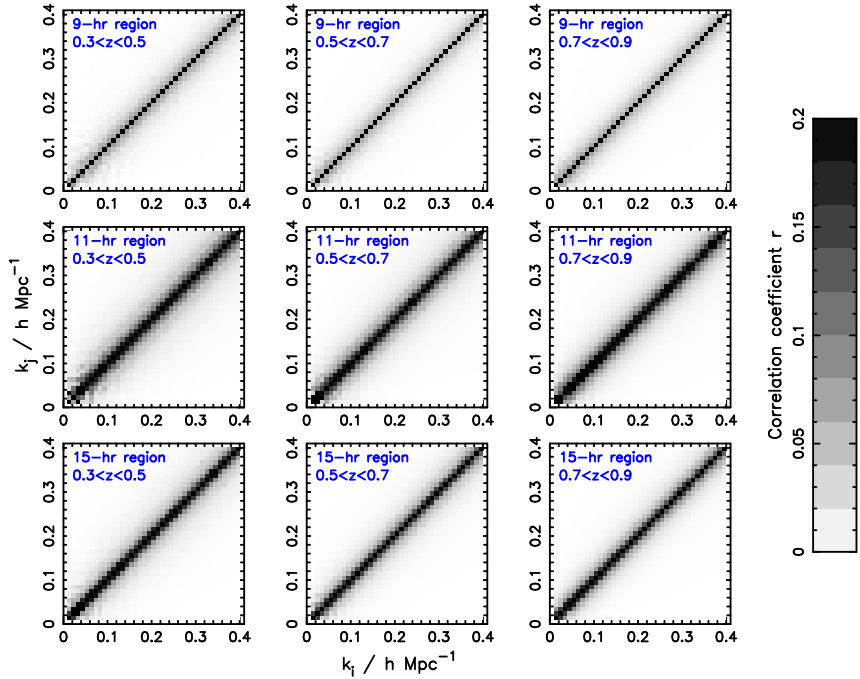}
\caption{Greyscale plot of the correlation coefficient $r$ of Equation
\ref{eqcoef} for each of the survey regions and redshift slices
analyzed in this paper, indicating the degree of covariance between
the power spectrum measurement in different Fourier bins.  Note the
choice of greyscale range such that the darkest shade corresponds to
$r = 0.2$.}
\label{figcov}
\end{figure*}

The corresponding nine covariance matrices $C_{ij}$ are plotted in
Figure \ref{figcov} as a correlation coefficient
\begin{equation}
r(i,j) = \frac{C_{ij}}{\sqrt{C_{ii} \, C_{jj}}}
\label{eqcoef}
\end{equation}
Figure \ref{figcov} demonstrates that the amplitude of the
off-diagonal elements of the covariance matrices is small (note the
choice of greyscale range).

We also measured power spectra in wavevector bins $(k_{\rm perp},
k_{\rm par})$ perpendicular and parallel to the line-of-sight,
respectively (we now use Fourier bins of width $\Delta k = 0.02 \, h$
Mpc$^{-1}$ in each direction to increase the signal-to-noise ratio in
each bin).  This 2D power spectrum allows us to recover the
redshift-space distortion parameters which produce an anisotropic
galaxy power spectrum.  Since in our analysis we orient the $x$-axis
parallel to the line-of-sight to the centre of each survey region, we
make the flat-sky approximation $k_{\rm perp} = \sqrt{k_y^2 + k_z^2}$,
$k_{\rm par} = |k_x|$ in this analysis.

\subsection{Systematic error study}

\begin{figure}
\center
\epsfig{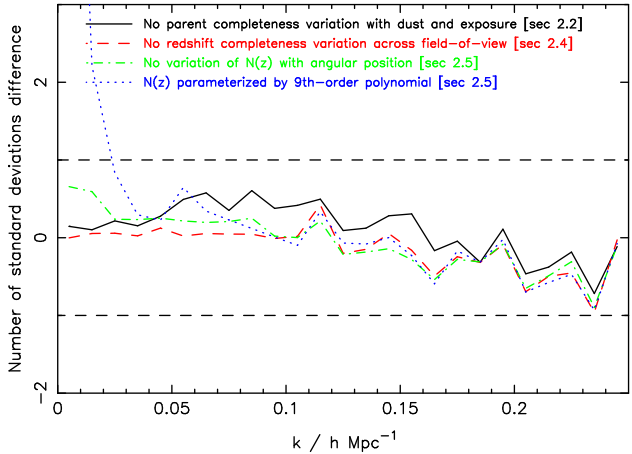}
\caption{Variations in the power spectrum estimate for the 9-hr
  region, in units of standard deviations in the fiducial power
  spectrum, obtained by changes in the methodology used to construct
  the selection function.}
\label{figpksys}
\end{figure}

We investigated the dependence of our power spectrum measurement on
potential systematic errors in the survey selection function.  In
order to do this we re-constructed four different selection functions
for the 9-hr region, analyzing the full redshift range $0.3 < z <
0.9$, with (extreme) variations in the method:
\begin{itemize}
\item We removed the variation of the completeness of the parent
  GALEX-SDSS sample with dust and GALEX exposure time described in
  Section \ref{secparent}, and instead assumed a constant density map.
\item We removed the variation of redshift completeness across the
  field-of-view of the 2dF spectrograph described in Section
  \ref{secfieldcomp}, and instead assumed a constant redshift
  completeness.
\item We removed the dependence of the survey redshift distribution on
  angular position described in Section \ref{secnz}, and instead used
  a position-independent radial selection function.
\item We parameterized the redshift distribution for each magnitude
  slice using a 9th-order polynomial rather than a sum of two Gaussian
  functions.
\end{itemize}
In Figure \ref{figpksys} we plot the difference between the resulting
power spectra measured for these four different selection functions
and the fiducial power spectrum, in units of the standard deviation of
the measurement at each scale.  We note that these (extreme)
variations in our understanding of the selection function typically
cause up to $\approx 0.5$-$\sigma$ shifts in the power spectrum
estimate.  The exception is the parameterization of the redshift
distribution, which causes large deviations in the very large-scale
($k < 0.03 \, h$ Mpc$^{-1}$) power spectrum.  For the example being
studied, we noted that the 9th-order polynomial fit was in fact
providing a poorer match than the double Gaussian function to the
observed redshift distributions, causing a spurious increase in
measured large-scale power.

We conclude that our power spectrum measurements are likely to be
robust against reasonable systematic variations in the selection
function methodology.  The selection function in Fourier space is compact
and the corrections are only significant on the largest scales.

\section{Cosmological parameter fits}
\label{secparfit}

In this Section we present some initial comparisons between our power
spectrum measurements and cosmological models.  Our aim is to
establish whether or not our measurements are consistent with the
predictions of the standard $\Lambda$CDM framework with parameters
determined by observations of the Cosmic Microwave Background
radiation.  Future papers will perform a more comprehensive analysis.

\subsection{Model power spectra}
\label{secpkmod}

We derived model matter power spectra $P_{\rm m}(k)$ as a function of
redshift using the {\tt CAMB} software package (Lewis, Challinor \&
Lasenby 2000) which is based on {\tt CMBFAST} (Seljak \& Zaldarriaga
1996), including corrections for non-linear growth of structure using
the fitting formulae of Smith et al.\ (2003) ({\tt halofit=1} in {\tt
  CAMB}).  Throughout this analysis (unless otherwise noted) we fix
values for the Hubble parameter $h = 0.72$, scalar index of spectral
fluctuations $n_{\rm s} = 0.96$ and normalization $\sigma_8 = 0.8$,
and vary only the matter density $\Omega_{\rm m}$ and the baryon
fraction $\Omega_{\rm b}/\Omega_{\rm m}$.  Our choices of parameter
values are consistent with recent observations of the Cosmic Microwave
Background radiation (Komatsu et al.\ 2009).

For this initial parameter fitting we assumed a linear bias factor $b$
for the galaxy population.  The assumption of linear bias will be
investigated in more detail in future studies through the use of mock
catalogues generated from dark matter simulations, analysis of the
galaxy bispectrum (e.g.\ Verde et al.\ 2002; Nishimichi et al.\ 2007),
and measurements of the halo occupation distribution of WiggleZ
galaxies (e.g.\ Zehavi et al.\ 2005; Blake, Collister \& Lahav 2008).
We note that in the current analysis the value of the bias factor is
significantly degenerate with the choice of $\sigma_8$ (such that $b
\propto 1/\sigma_8$).

We modified this real-space matter power spectrum to include
redshift-space distortions, which we characterized in the standard
manner using two parameters.  The first parameter, $f$, models the
effect of large-scale coherent infall velocities.  In linear theory
for the growth of fluctuations, $f$ is related to the linear growth
factor $D(a)$ at scale factor $a$ by $f = d\ln{D}/d\ln{a}$.  This
functional form is well-approximated by $f = \Omega_{\rm m}(z)^\gamma$
where $\Omega_{\rm m}(z)$ is the matter density measured by an
observer at redshift $z$, and $\gamma \approx 0.55$ for a standard
$\Lambda$CDM cosmology.  Coherent velocities can also be parameterized
by $\beta = f/b$, introducing a dependence on galaxy type via the bias
factor.  The second parameter, $\sigma_v$ (in units of $h$ km
s$^{-1}$), describes small-scale random virialized velocities, for
which we assumed an exponential pairwise distribution.  The overall
effect on the galaxy power spectrum $P_{\rm gal}$ is to induce an
anisotropic distortion dependent on the angle $\theta$ of the
wavevector to the line-of-sight, parameterized by $\mu =
\cos{\theta}$:
\begin{equation}
P_{\rm gal}(k,\mu) = b^2 \, P_{\rm m}(k) \, \frac{(1 + f \mu^2/b)^2}{1
  + (k \, H_0 \, \sigma_v \, \mu)^2}
\label{eqpkmod}
\end{equation}

\subsection{Fits to the 2D power spectra in redshift slices}

We fitted these models to the nine independent 2D power spectra split
into tangential and radial components, in order to determine the
redshift-space distortion parameters and test the $\Lambda$CDM
self-consistency relation $\beta \approx \Omega_{\rm m}(z)^{0.55}/b$
as a function of redshift.  In this initial analysis we fixed the
matter density parameter $\Omega_{\rm m} = 0.3$ and baryon fraction
$\Omega_{\rm b}/\Omega_{\rm m} = 0.166$ (similar to the parameters
listed in Komatsu et al.\ (2009) that provide the best fit to the
5-year data set of the Wilkinson Microwave Anisotropy Probe (WMAP); we
fix $\Omega_{\rm m} = 0.3$ to match the value assumed for the
distance-redshift relation in our power spectrum analysis).  We then
varied the redshift-space parameters $f$ and $\sigma_v$ and the bias
factor $b$, convolving each model power spectrum with the survey
selection function for each region and minimizing the chi-squared
statistic calculated using the full covariance matrices.  For each
redshift slice we combined the results from the three survey regions.
The best-fitting models produce a good fit to the measured power
spectrum, as evidenced by the values of the chi-squared statistic in
the three redshift slices: $\chi^2 = (1113.9, 1229.6, 1069.2)$ for
1137 degrees of freedom.

\begin{figure}
\center
\epsfig{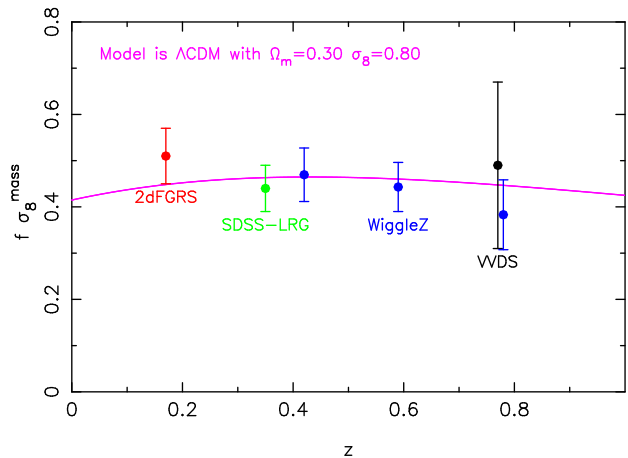}
\caption{Measurements of the growth rate of structure, $f$, obtained
  in three redshift slices by fitting WiggleZ survey data in three
  survey regions.  The values of $f$ are weighted by $\sigma_{8,{\rm
      mass}}(z)$ as discussed by Song \& Percival (2009).  The
  measurements are compared to results previously published for the
  2dFGRS, SDSS LRG, and VIRMOS-VLT Deep Survey (VVDS) samples, as
  collected by Song \& Percival.  A prediction for the $\Lambda$CDM
  cosmological model with $\Omega_{\rm m} = 0.3$ and $\sigma_8 = 0.8$
  is overplotted.}
\label{figgrowth}
\end{figure}

The measurements of the growth rates at each effective redshift
$z_{\rm eff} = (0.42, 0.59, 0.78)$ are $f = (0.73 \pm 0.09, 0.75 \pm
0.09, 0.71 \pm 0.14)$, marginalizing over the parameters $\sigma_v$
and $b$, where 1-$\sigma$ error ranges are indicated.  These should be
compared to the $\Lambda$CDM prediction for $\Omega_{\rm m}(0) = 0.3$:
$f = (0.72,0.78,0.83)$.  We find that the model successfully passes
this test of self-consistency.  Our measurements of the growth rate
are compared with the model in Figure \ref{figgrowth} by plotting the
quantity $f \times \sigma_{8,{\rm mass}}(z)$ versus redshift.
Previously-published results are also indicated, as summarized by Song
\& Percival (2009).  The WiggleZ Survey data provides a growth
measurement across the redshift range $0.4 < z < 0.8$ with a precision
similar to previous work at $z < 0.4$.  We highlight in particular the
$20\%$ accuracy of our measurement at $z = 0.78$.  A future study will
perform a more detailed analysis of these results.

The fitted values of the pairwise velocity dispersion in each redshift
slice are $\sigma_v = (354 \pm 41, 294 \pm 31, 216 \pm 58) \, h$ km
s$^{-1}$ (marginalizing over $f$ and $b$).  We find evidence that the
pairwise small-scale velocity dispersion of WiggleZ galaxies
systematically decreases with increasing redshift.  The measured bias
factors for each redshift slice are $b = (0.93 \pm 0.03, 1.08 \pm
0.03, 1.20 \pm 0.06)$ (marginalizing over $f$ and $\sigma_v$).  We can
compare these galaxy bias measurements to those deduced from the
WiggleZ survey small-scale correlation function measurements: for
three redshift slices similar to those analyzed here, Blake et
al.\ (2009) obtained $b = (1.01, 1.27, 1.27)$ (although assuming a
higher value of $\sigma_8 = 0.9$).  The main cause of this difference
is the superior determination of $\beta$ in the current analysis.
Other issues wth this comparison include: (i) galaxy bias is a
scale-dependent function, (ii) small-scale pairwise velocities were
not been modelled in the Blake et al.\ (2009) analysis, (iii) the
power-law correlation function model assumed in Blake et al.\ (2009)
breaks down at large scales.

\subsection{Stacked 1D power spectra}

Using these model fits we can combine the nine 1D power spectrum
measurements in the different regions and redshift slices into a
single ``stacked'' survey 1D power spectrum using inverse-variance
weighting.  The difficulty with this step is that the power spectrum
amplitude in each region is modulated by a different level of
convolution with the selection function, as illustrated by Figure
\ref{figpkreg}.  In addition, the shape of the power spectrum varies
with redshift in accordance with the differing redshift-space
distortion parameters and galaxy bias factors.

\begin{figure*}
\center
\epsfig{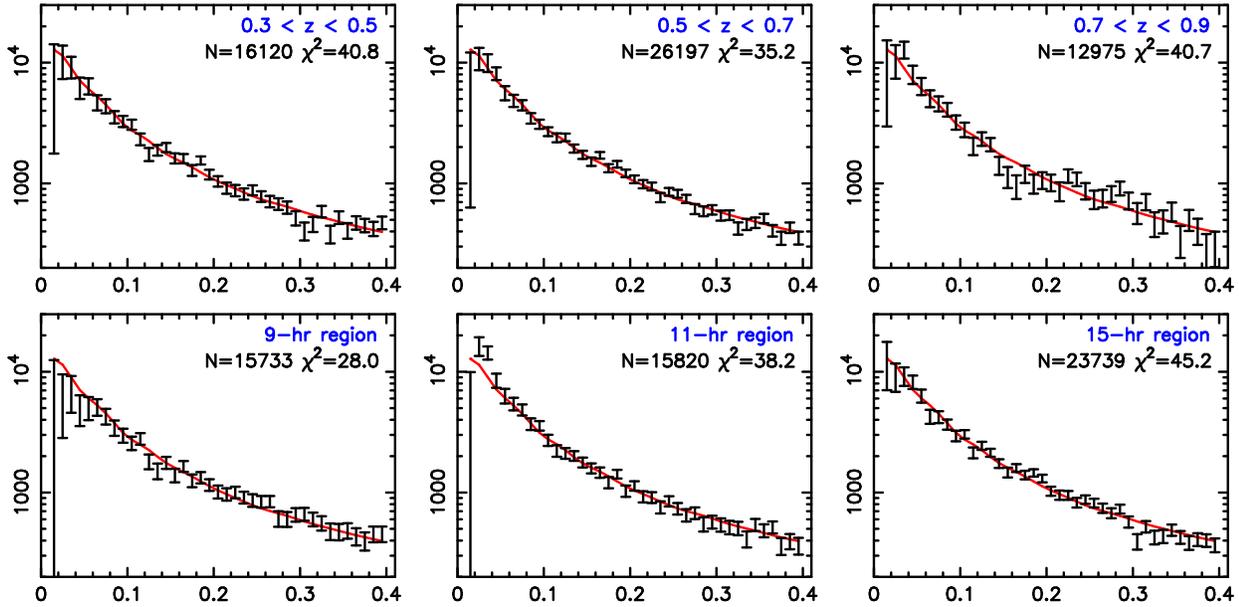}
\caption{The power spectra determined by combining the measurements in
  the different survey regions (for each redshift slice) and different
  redshift slices (for each survey region).  Before combination, we
  adjust the shapes of the power spectra to allow for the differing
  degrees of convolution with the selection function and the differing
  redshift-space effects.  The solid line in the panel has been
  generated by combining the best-fitting model power spectra in a
  similar manner.  The $x$-axis and $y$-axis of each panel are Fourier
  wavescale $k$ in units of $h$ Mpc$^{-1}$ and power spectrum
  amplitude $P(k)$ in units of $h^{-3}$ Mpc$^3$, respectively.}
\label{figpkstack}
\end{figure*}

\begin{figure*}
\center
\epsfig{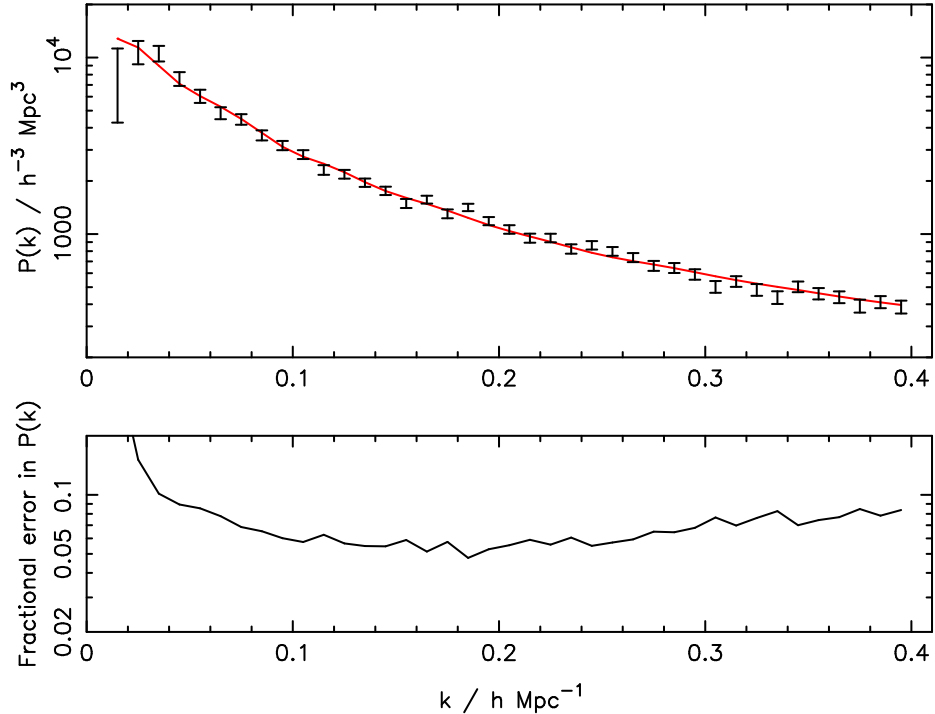}
\caption{{\it Upper panel:} The power spectrum determined by combining
  the measurements in the different survey regions and redshift slices
  analyzed in this paper using inverse-variance weighting.  Before
  combination, we adjust the shapes of the power spectra to allow for
  the differing degrees of convolution with the selection function and
  the differing redshift-space effects.  The solid line in the panel
  has been generated by combining the best-fitting model power spectra
  in a similar manner.  {\it Lower panel:} The fractional error in the
  combined power spectrum as a function of Fourier wavenumber $k$.}
\label{figpkcomb}
\end{figure*}

\begin{figure*}
\center
\epsfig{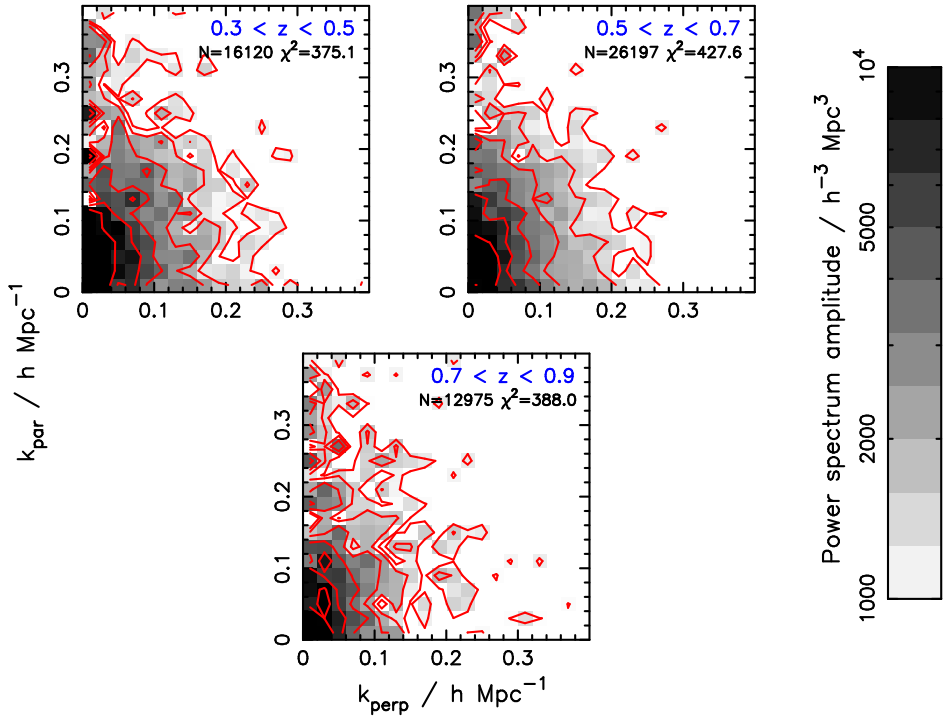}
\caption{The 2D galaxy power spectra for each of the redshift slices
  analyzed in this paper, obtained by combining measurements in
  different survey regions, and plotted as a function of wavevectors
  $(k_{\rm perp}, k_{\rm par})$ tangential and radial to the
  line-of-sight, respectively.  The function is represented using both
  greyscale and contours; the contour levels are logarithmically
  spaced between $P = 1000$ and $10{,}000 \, h^{-3}$ Mpc$^3$.  The
  non-circularity of the contours encodes the imprint of large-scale
  galaxy peculiar velocities, as discussed in the text.}
\label{figpk2red}
\end{figure*}

Therefore, before combining the results in the different regions and
redshift slices, we performed an approximate ``de-convolution'' of
amplitude by multiplying the power spectra by a correction factor
equal to the scale-dependent ratio of the convolved and unconvolved
model power spectra, i.e.\ the ratio of the solid and dashed curves
plotted in Figure \ref{figpkreg}.  (In detail we did not use the model
power spectra to generate these corrections but a ``reference'' power
spectrum in which the baryon oscillation features have been smoothed
out, i.e.\ the ``no-wiggles'' power spectrum of Eisenstein \& Hu
(1998).  We preferred to correct the amplitude using a wiggle-free
power spectrum to avoid spuriously introducing apparent baryon
oscillations into the combined power spectrum).  We additionally
corrected the measurement in each Fourier bin by the angle-averaged
ratio of the redshift-space galaxy power spectrum (using the
best-fitting values of $\beta$ and $\sigma_v$ for each redshift slice)
to the real-space matter power spectrum at redshift $z=0.6$.
Following these corrections, the nine power spectra ``line up'' with
consistent shape and amplitude, and can be combined.  In Figure
\ref{figpkstack} we show separate combinations of power spectra across
survey regions in each redshift slice, and across redshift slices in
each survey region.  The combination of all nine power spectra is
presented in Figure \ref{figpkcomb}; the lower panel indicates the
fractional accuracy of the measurement, which approaches $5\%$ in
Fourier bins of width $\Delta k = 0.01 \, h$ Mpc$^{-1}$.  The scale
dependence of the fractional accuracy is determined by a balance
between the increasing number of Fourier modes contributing to each
successive bin (tending to decrease the error), and the increasing
importance with $k$ of shot noise relative to cosmic variance (tending
to increase the error).  Finally, in Figure \ref{figpk2red} we display
the 2D power spectra for each redshift slice, combining different
regions in the manner described above.

\subsection{Fits for matter and baryon densities}

As a final exercise we fitted the measured power spectra for the
cosmological matter and baryon densities (parameterized by
$\Omega_{\rm m}$ and $f_{\rm b} = \Omega_{\rm b}/\Omega_{\rm m}$).  We
tried three different initial approaches to this analysis, with
increasing degrees of sophistication:

\begin{itemize}

\item First, we generated unconvolved, real-space, non-linear power
  spectra and fitted the combined survey data points plotted in the
  upper panel of Figure \ref{figpkcomb} by varying $\Omega_{\rm m}$,
  $f_{\rm b}$ and the bias factor $b$.

\item Second, we fitted the convolved, redshift-space 1D power spectra
  in each of the survey regions and redshift slices, fixing the
  redshift-space distortion parameters $\beta$ and $\sigma_v$ at the
  best-fitting measurements using the 2D power spectrum fits at each
  redshift and varying $\Omega_{\rm m}$, $f_{\rm b}$ and $b$.

\item Third, we fitted the convolved, redshift-space 2D power spectra,
  fixing the value of $\sigma_v$ at the best-fitting values.  We
  varied $\Omega_{\rm m}$, $f_{\rm b}$ and the bias factor $b$,
  determining the value $\beta = \Omega_{\rm m}(z)^{0.55}/b$ in each
  case (assuming the $\Lambda$CDM model).  For each different value of
  $\Omega_{\rm m}$, we scaled $\sigma_8$ assuming the ``WMAP
  normalization'' (Komatsu 2009, Section 5.5) and also scaled the
  geometry of the survey box in accordance with each trial cosmology:
  given a fiducial co-moving distance $D_{\rm fid}$ and Hubble
  parameter $H_{\rm fid}$ for a redshift slice calculated using
  $\Omega_{\rm m} = 0.3$, and also given trial values $D$, $H$
  corresponding to a different value of $\Omega_{\rm m}$, the width of
  Fourier bins tangential and radial to the line-of-sight scales as
  $D/D_{\rm fid}$ and $H_{\rm fid}/H$, respectively; furthermore the
  volume of the box (hence amplitude of the measured power spectrum)
  scales as $(D/D_{\rm fid})^2 \times (H_{\rm fid}/H)$.

\end{itemize}

\begin{figure}
\center
\epsfig{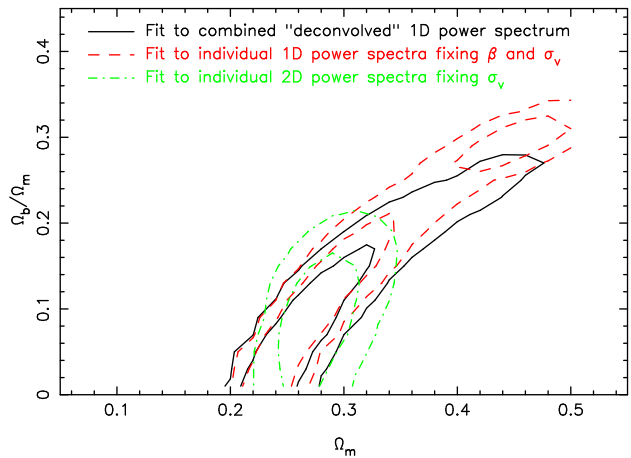}
\caption{Probability contours of $\Omega_{\rm m}$ and $\Omega_{\rm
    b}/\Omega_{\rm m}$ fitting to WiggleZ survey power spectra using
  the three different approaches described in the text.  The inner and
  outer contours for each set enclose $68\%$ and $95\%$ of the
  likelihood, respectively.}
\label{figomfbprob}
\end{figure}

Figure \ref{figomfbprob} displays the probability contours in the 2D
space of $\Omega_{\rm m}$ and $f_{\rm b} = \Omega_{\rm b}/\Omega_{\rm
  m}$, marginalizing over the bias factor, for each of the three
approaches described above.  The different methods produce broadly
consistent results, and the best-fitting values are in good agreement
with those determined from the latest measurements of the temperature
fluctuations of the Cosmic Microwave Background radiation (Komatsu et
al.\ 2009).

The probability contours in Figure \ref{figomfbprob} display evidence
of the well-known degeneracy between $\Omega_{\rm m}$ and $\Omega_{\rm
  b}/\Omega_{\rm m}$ in the determination of the overall shape of the
matter power spectrum, which is exacerbated by the fact that we cannot
yet detect the imprint of the baryon acoustic oscillations in our
power spectrum measurement.  This degeneracy will be broken as the
WiggleZ Survey progresses.

\section{Conclusions}
\label{secconc}

In this paper we have described our method of determining the
selection function of the WiggleZ Dark Energy Survey, and have
presented the current measurement of the large-scale galaxy power
spectrum using $56{,}159$ redshifts of bright emission-line galaxies
spanning redshifts $0.3 < z < 0.9$.  This sample constitutes approximately
$25\%$ of the final WiggleZ survey.  We have quantified and
categorized the redshift blunder rate and determined its effect on the
power spectrum measurement via analytical calculations and detailed
simulations.  We conclude that:
\begin{itemize}
\item The selection function of the WiggleZ survey is complicated by
  the proximity of the faint magnitude threshold to the completeness
  limit of the input catalogues, in particular for the GALEX UV data.
  We quantified the incompleteness in the parent target catalogue as a
  function of GALEX exposure time and Galactic extinction via fitting
  formulae.
\item We adopted a Monte Carlo technique to determine the relative
  completeness of the spectroscopic follow-up at any position.  This
  technique allows for the complex overlapping of survey pointings and
  for the systematic variation of redshift completeness across the
  2-degree field-of-view of the instrument.  We also allowed for the
  magnitude prioritization of the spectroscopic follow-up which
  results in a position-dependent galaxy redshift distribution.
\item The WiggleZ survey contains redshift blunders resulting from
  emission-line confusion (most significantly, [OIII], H$\beta$ and
  H$\alpha$ mis-identified as [OII]) and from sky emission lines
  mis-identified as [OII].  The overall blunder rate is about $5\%$.
  The effect of the redshift blunders on the power spectrum
  measurement is well-approximated as a constant reduction in
  amplitude for scales $k > 0.05 \, h$ Mpc$^{-1}$ combined with an
  enhanced level of reduction for large scales $k < 0.05 \, h$
  Mpc$^{-1}$.
\item We measured 1D (angle-averaged) and 2D (binned in tangential and
  radial modes) galaxy power spectra for nine independent survey
  regions and redshift slices using the method of Feldman, Kaiser \&
  Peacock (1994).  The 1D power spectrum for the whole sample,
  combining these measurements, has a fractional accuracy of about
  $5\%$ in Fourier bins of width $\Delta k = 0.01 \, h$ Mpc$^{-1}$.
  The 2D power spectra show the expected anisotropic signatures of
  redshift-space distortions due to large-scale coherent infall and
  small-scale virialized motions.
\item The power spectrum data are well-described by a model power
  spectrum with matter and baryon densities consistent with those
  determined from observations of the Cosmic Microwave Background
  radiation.  The model includes non-linear corrections,
  redshift-space distortions and a linear galaxy bias factor.
\item The 2D power spectra allow us to measure the growth rate of
  structure across the redshift range $0.4 < z < 0.8$.  We obtain
  results similar in precision to previous determinations at $z <
  0.4$, including a measurement at $z=0.78$ with $20\%$ accuracy.
\end{itemize}
Future studies will present full cosmological parameter fits and
combinations of these results with other datasets, including
implications for the growth of cosmic structure and Gaussianity of the
initial conditions, and extend these analyses to the final WiggleZ
survey catalogues.

\section*{Acknowledgments}

We thank an anonymous referee for useful comments on the submitted
version of this paper.

We acknowledge financial support from the Australian Research Council
through Discovery Project grants funding the positions of SB, MP, GP
and TD.  SMC acknowledges the support of the Australian Research
Council through a QEII Fellowship.  MJD thanks the Gregg Thompson Dark
Energy Travel Fund for financial support.

GALEX (the Galaxy Evolution Explorer) is a NASA Small Explorer,
launched in April 2003.  We gratefully acknowledge NASA's support for
construction, operation and science analysis for the GALEX mission,
developed in co-operation with the Centre National d'Etudes Spatiales
of France and the Korean Ministry of Science and Technology.

Finally, the WiggleZ survey would not be possible without the
dedicated work of the staff of the Anglo-Australian Observatory in the
development and support of the AAOmega spectrograph, and the running
of the AAT.

\end{document}